\documentclass[twocolumn]{aastex62}

\usepackage{xspace}
\usepackage{natbib}
\usepackage{verbatim}
\usepackage{graphicx,epstopdf}
\usepackage{longtable}
\usepackage{fancyref}
\usepackage{hyperref}
\usepackage{breakurl}
\usepackage{amsmath}
\usepackage{graphicx} 
\usepackage{epstopdf}
\usepackage{graphicx}
\usepackage{appendix}

\begin{document}

\author{Jun-Jie Jin}
\affiliation{Key Laboratory of Optical Astronomy, National Astronomical Observatories, Chinese
	Academy of Sciences, Beijing 100012, P.R. China;\ E-mail: hwu@bao.ac.cn}
\affiliation{School of Astronomy and Space Science University of Chinese Academy of Sciences, Beijing 100012, P.R. China;\ E-mail: jjjin@bao.ac.cn}

\author{Yi-Nan Zhu}
\affiliation{School of Physics and Astronomy, Sun Yat-sen University, Zhuhai 519082, P.R.China; \ E-mail: zhuyn9@mail.sysu.edu.cn}

\author{Hong Wu}
\affiliation{Key Laboratory of Optical Astronomy, National Astronomical Observatories, Chinese
	Academy of Sciences, Beijing 100012, P.R. China;\ E-mail: hwu@bao.ac.cn}
\affiliation{School of Astronomy and Space Science University of Chinese Academy of Sciences, Beijing 100012, P.R. China;\ E-mail: jjjin@bao.ac.cn}
\author{Feng-Jie Lei}
\affiliation{Key Laboratory of Optical Astronomy, National Astronomical Observatories, Chinese
	Academy of Sciences, Beijing 100012, P.R. China;\ E-mail: hwu@bao.ac.cn}
\affiliation{School of Astronomy and Space Science University of Chinese Academy of Sciences, Beijing 100012, P.R. China;\ E-mail: jjjin@bao.ac.cn}

\author{Chen Cao}
\affiliation{School of Space Science and Physics, Shandong University, Weihai, Weihai, Shandong 264209, P.R.China}

\author{Xian-Min Meng}
\affiliation{Key Laboratory of Optical Astronomy, National Astronomical Observatories, Chinese
	Academy of Sciences, Beijing 100012, P.R. China;\ E-mail: hwu@bao.ac.cn}

\author{Zhi-Min Zhou}
\affiliation{Key Laboratory of Optical Astronomy, National Astronomical Observatories, Chinese
	Academy of Sciences, Beijing 100012, P.R. China;\ E-mail: hwu@bao.ac.cn}

\author{Man I Lam}
\affiliation{Leibniz-Institut f$\ddot{u}$r Astrophsik Potsdam (AIP), An der Sternwarte 16.14482 Potsdam, Germany}


\title{An $H\alpha$ Imaging Survey of the all (Ultra-)Luminous Infrared Galaxies at $Dec. \ge -30^{\circ}$ in the GOALS Sample }

\begin{abstract}
This paper presents the result of $H\alpha$ imaging for luminous infrared galaxies (LIRGs) and ultra-luminous infrared galaxies (ULIRGs). \textbf{It is } a complete subsample of Great Observatories All-sky LIRG Survery (GOALS) with $Dec. \ge -30^{\circ}$, and consists 148 galaxies with $log(L_{IR}/L_{\sun}) \ge 11.0$. All the $H\alpha$ images were carried out using the 2.16-m telescope at the Xinglong Station of the National Astronomy Observatories, Chinese Academy of Sciences (NAOC), during the year from 2006 to 2009. We obtained pure $H\alpha$ luminosity for each galaxy and corrected the luminosity for $[NII]$ emission, filter transmission and extinction. We also classified these galaxies based on their morphology and interaction. We found that the distribution of star-forming regions in these galaxies is related to this classification. As the merging process advanced, these galaxies tend to have a more compact distribution of star-forming region, higher $L_{IR}$ and warmer IR-color ($f_{60}/f_{100}$). These results imply that the degree of dynamical disturbance plays an important role in determining the distribution of star-forming region.

\end{abstract}


\section{Introduction} \label{sec:intro}
Luminous infrared galaxies (LIRGs) and ultraluminous infrared galaxies (ULIRGs) are galaxies with infrared luminosity $log(L_{IR}/L_{\sun})\ge 10^{11}$ and $log(L_{IR}/L_{\sun}) \ge 10^{12}$, respectively. The research for them began after the successful launch of Infrared Astronomical Satellite (IRAS) in 1983 \citep{1972ApJ...178..623T,1977ARA&A..15..437T,1988ApJ...325...74S}. These galaxies often show signs of tidal interaction and merger, which lead to the extreme nuclear activity and strong starburst (SB) \citep{1995PhDT........11K,1995ApJS...98..171V,1991ApJ...370L..65B,1992ARA&A..30..705B,1996ApJ...471..115B,1991MNRAS.252..593Z,1996ARA&A..34..749S,1998A&AS..127..521W,1998A&AS..132..181W,1999A&A...349..735Z,2001AJ....122...63C,2002ApJ...564..196X,2005MNRAS.361..776S,2006ChJAA...6..197C,2006ApJ...649..722W,2015RAA....15.1424L}.  

 An important question regarding the nature of (U-)LIRGs is what phase do they play in the general evolution of galaxies?  Many statistical studies showed that the interaction between galaxies can enhance star formation activity \citep{1987AJ.....93.1011K}. And the ULIRGs are associated with the interaction \citep{1991MNRAS.252..593Z,1988ApJ...325...74S,1998A&AS..127..521W,1998A&AS..132..181W} as well as active galaxy nuclear (AGN) activities \citep{1998A&AS..127..521W,1998A&AS..132..181W}. Theoretical and observational works support an evolution scenario where two gas-rich spiral galaxies merge first and drive material from galaxies disk toward the merger center, triggering star-formation activity before dust-enshrouded AGN in the circumnuclear region\citep{1988ApJ...325...74S,1992ARA&A..30..705B,2008ApJS..175..356H,2018ApJ...864...32J}.

Former works showed that the degree of interaction has a great influence on the star formation process in ULIRGs.  \cite{2004AJ....127..736H} performed an $H\alpha$ imaging observation for 22 LIRGs and found that the distribution of star-formation region is strongly related to the property of galaxy interactions.  Many works also showed that the interaction/merger rate increases with IR luminosity, and nearly all (U-)LIRGs show signs of interaction/merger events \citep{1998A&AS..132..181W,2002ApJS..143..277K,2002ApJS..143..315V,1991MNRAS.252..593Z}. \cite{2016ApJ...825..128L} presented an analysis of morphologies and molecular fraction (MGFs) for 65 LIRGs in GOALS sample. They  found that the mean MGF for non-interacting LIRGs is much less than that of intermediate stage in major merger LIRGs. 

However, the diversities of star-forming and morphological properties in (U-)LIRGs are not fully understood. For example, \cite{1996ARA&A..34..749S} suggested that $20 \sim 30\%$ of LIRGs with $10^{11}L_{\sun} < L_{IR} <10^{12}L_{\sun}$ are apparently single galaxies.  \cite{1999AJ....118.2625R} presented a mid-IR spectroscopic survey for 62 ULIRGs and found there is no correlation between merging process and infrared luminosity ($L_{IR}$) which is contrary to conventional ULIRGs evolutionary scenario. The numerical simulation showed that the  minor merger between gas-rich disks and less massive dwarf galaxies can also produce nuclear starbursts \citep{1995ApJ...448...41H}.
 
 By using the luminosity ($H\alpha$, UV, IR) related to young massive stars, we can get the information of star-formation. Among them, the $H\alpha$ emission is proportional to the number of ionizing photos which are produced by young stars with age less than $\sim$10Myr and mass higher than $17M _{\sun}$ \citep{2016MNRAS.455.1807W}. Therefore, the $H\alpha$ emission directly traces the presence of recently formed massive stars. Although former studies yield many interesting results, their samples are not enough \citep{2004AJ....127..736H} or biased on one type of galaxies \citep{2016ApJ...822...45T,1996AJ....112.1903Y,2013ApJ...768..102K}.  Great Observatories All-sky LIRG Survey (GOALS)  \footnote {http://goals.ipac.caltech.edu} is a subset of  the IRAS Revised Bright Galaxy Sample  \citep [RBGS;][]{2003AJ....126.1607S}. In order to better understand the property of (U-)LIRGs, we have undertaken an $H\alpha$ imaging survey of 148 (U-)LIRGs selected from GOALS which will help us to study the star-formation in (U-)LIRGs.

In this paper, we present initial result for our $H\alpha$ imaging survey on 148 GOALS sample galaxies. The layout of this paper is as follow: In Section 2, the sample selection and observation are summarized. In Section 3, we describe the data reduction. And in Section 4, we present the main results of this survey which include the $H\alpha$ catalog and reduced $H\alpha$ images. The results based on morphological classification are presented in Section 5 and the discussion is provided in Section 6. At last, a summary of the paper is presented in Section 7. Throughout this paper, we adopt the cosmology $H_{0}=75$ $km$ $s^{-1}$ $Mpc^{-1}$ and a flat universe where $\Omega_{M}=0.3$ and $\Omega_{\Lambda}=0.7$.

\section{Sample and Observations}
\subsection{Sample}

\cite{2003AJ....126.1607S} provided a complete flux-limited extragalactic sample (RBGS: the IRAS Revised Bright Galaxy Sample) with $60 \mu m$ flux densities greater than 5.24 Jy and $\lvert b \lvert > 5^{\circ}$. \cite{2009PASP..121..559A} constructed the GOALS sample from RBGS, which includes 181 LIRGs and 21 ULIRGs in the local universe ($z < 0.088$). This sample combines data from NASA's \emph{Spizer Space Telescope}, \emph{Chandra X-Ray Observatory}, \emph{Hubble Space Telescope (HST)}, and \emph{Galaxy Evolution Explorer (GALEX)} observatories, together with ground-base data \citep{2009PASP..121..559A}. \cite{2017ApJS..229...25C} have presented broadband \emph{Herschel} imaging for entire GOALS sample for all six \emph{Herschel} bands (PACS bands: 70, 100, 160 $\mu m$; SPIRE bands: 250, 350, 500 $\mu m$). This sample span a wide range of nuclear spectral types and interaction stages which provide an unbiased picture of the (U-)LIRGs in local universe. Out of the original list of 202 sources in the GOALS sample, two objects are omitted. One is NGC 5010, whose $L_{IR}$ drops significantly below the LIRG threshold ($log(L_{IR}/L_{\sun}) = 11.0$) due to a revised redshift. The other is IRAS 05223-1908, which is proved as a young stellar object \citep{2017ApJS..229...25C}.

Our (U-)LIRGs sample is a subset of the GOALS sample. Considering the observatory site (around 40 degrees north latitude), our sample include 148 objects with $Dec. \ge -30^{\circ}$

The full sample along with their basic properties is listed in Table~\ref{table:RBGS} . The detailed measurements can be found in \cite{2009PASP..121..559A}. 
Given the poor IRAS resolution, sometimes there may be a pair or multiple galaxies within IRAS 3 $\sigma$ uncertainty ellipse. In such case, we visually examined the region covered by the IRAS 3 $\sigma$ uncertainty ellipse in the continuum-subtracted $H\alpha$ images (achieved after data reduction  ) and include all obvious  objects within this region. The different counterparts in the system are represented by N (north), S (south), E (east) or W (west) in Table~\ref{table:RBGS}.  There are 10 such galaxy pairs in our sample and each contains two counterparts. So, in the end, there are total 158 sources in our sample.

\begin{figure*}[!htb]
	\begin{center}
		\includegraphics[angle=0,scale=0.4,keepaspectratio=flase]{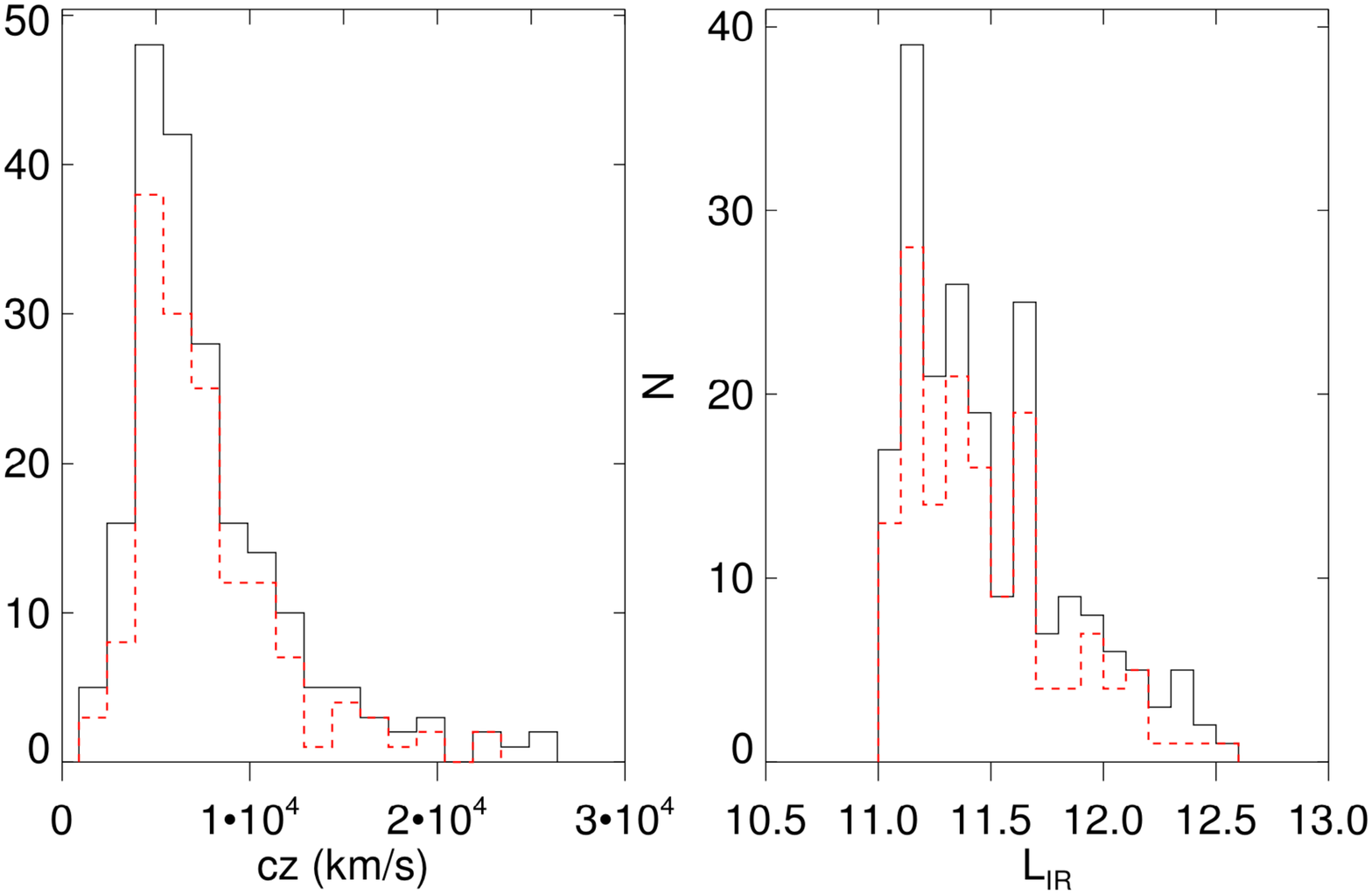}
		\caption{The distribution of cz and $L_{IR}$  of GOALS and our sample. The black solid lines represent the GOALS and the red  dashed lines represent our sample. \label{fig:GOALS}}
	\end{center}
\end{figure*}

Figure ~\ref{fig:GOALS} shows the sample distribution of $L_{IR}$ and recession velocity (cz).  The black solid line represents the GOALS sample and the red dashed line represents our sample. It is clear that two sample have the similar distribution and there is no significant differences between them. The cz concentrates in the range of 4000 $km$ $s^{-1}$to 7000 $km$ $s^{-1}$ corresponding to the redshift from 0.013 to 0.023. Left panel shows the $L_{IR}$ distribution. The number of galaxies decrease rapidly as the $L_{IR}$ increases.  Most galaxies in our sample are LIRGs with $log(L_{IR}/L_{\sun})$ lower than 12.0.

\subsection{Observations}
The observation were carried out on the 2.16-m telescope at the Xinglong Station of the National Astronomical Observatories, CAS. All the galaxies in our sample were taken in dark night between 2006 February and 2009 June. We used BAO Faint Object Spectrograph and Camera (BFOSC), which had $2048 \times 2048$ $pixel^{2}$ with the pixel scale of 0.30 $arcsec$ and has a field of view (FOV) of $11. \times 11.$ $arcmin^{2}$. We adopted a readout noise of 8.6 $e^{-} pixel^{-}$ with an average gain of 1.65 $e^{-}  ADU^{-}$ during the observation. The latest description of updated parameters for BFOSC can be seen in \cite{2016PASP..128k5005F}.

To obtain the distribution of star-formation region, we observed with both the narrow $H\alpha$ filter which cover the shifted $H\alpha$ emission at the velocity of the target galaxy and the broad R-band filter which used to determine the nearby continuum level. There are a series of narrow band $H\alpha$ filters, whose center wavelength ranges from 6563 $\AA$ to 7052 $\AA$ with the FWHM about 55 $\AA$. The detail description of $H\alpha$ filter can be seen in \cite{2018ApJS..235...18L}. The effective wavelength $\lambda_{eff}$ of the broad R-band filter is 6407$\AA$. 

For each object, the typically integral time is about 600s in R-band and 3600s in $H\alpha$. Table~\ref{table:RBGS} lists the $H\alpha$ filters for each sources used in observation together with all the other observation information.

\section{Data Reduction}

\subsection{Image Preprocessing}
After the observation, we checked the quality of the images by naked eyes. The subsequent data reduction were performed using IRAF, including overscan subtraction, bias subtraction, flat-field correction. The cosmic-rays were identified and removed using L.A.Cosmic \citep{2001PASP..113.1420V}. Then, the celestial coordinate was added to each image using $Astrometry.net$\footnote {http:\/\/nova.astrometry.net} and the bad columns were replaced with a linear fit of surrounding pixels .

\begin{figure*}[!htb]
	\begin{center}
		\includegraphics[angle=0,scale=0.6,keepaspectratio=flase]{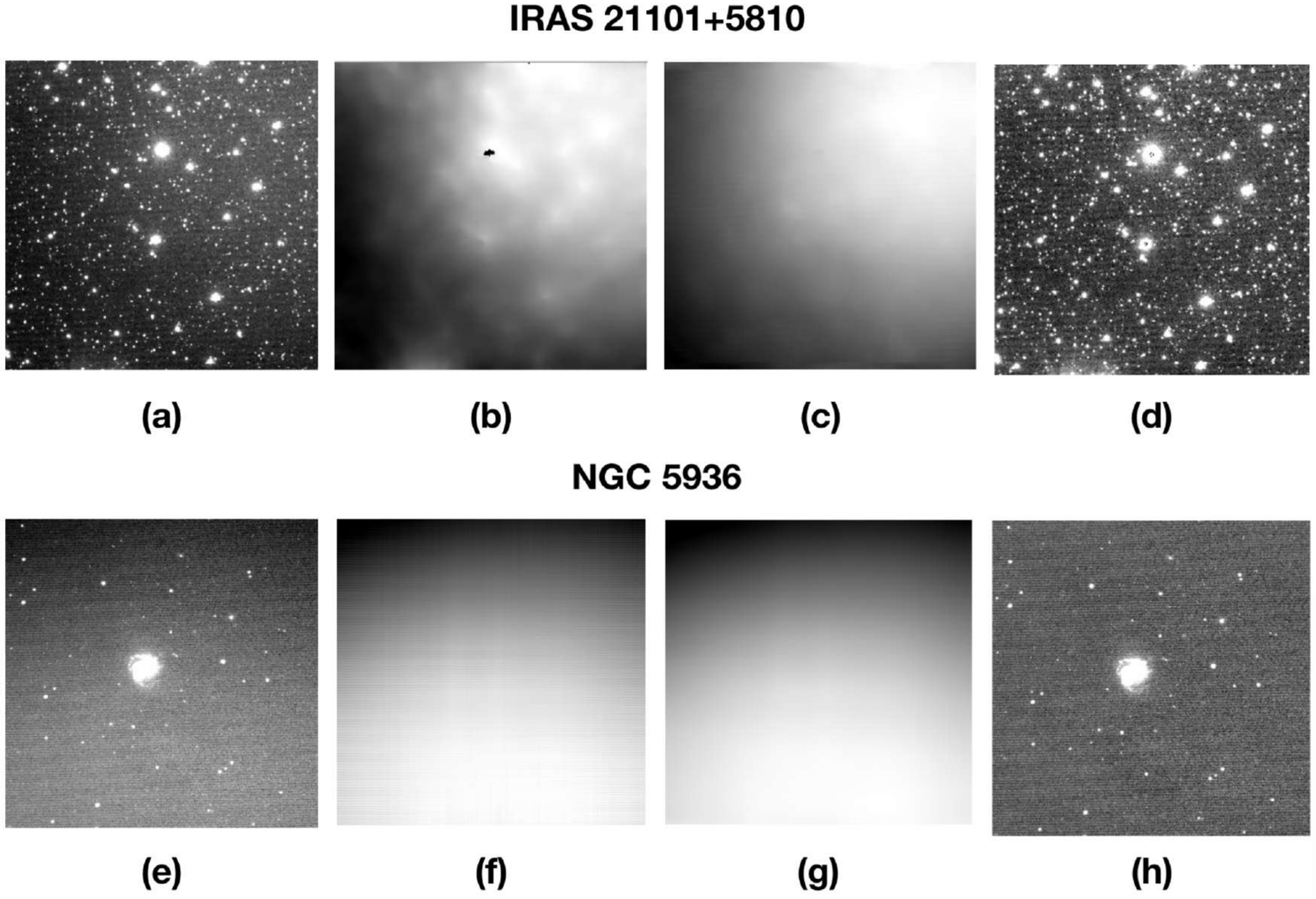}
		\caption{An example of sky-subtraction. Top panel: sky subtraction process for LIRGs IRAS 21101+5810 (method-1).  The four panels show the: (a) original image; (b) the sky background directly got from original image; (c) average stacking sky background; (d) sky-subtracted image.  Bottom panel: sky subtraction process for NGC 5936 (method-2).  The four panels show the: (e) original image; (f) the average sky image of row-fitted and column-fitted sky image; (g) the smoothed sky background image ; (h) sky-subtracted image.\label{fig:sub_SKY}}
	\end{center}
\end{figure*}

The next step is sky subtraction.  The most critical step is the sky background construction.  Firstly, Sextractor was employed to detect objects. Before detecting, we produced a gaussian smoothed image by convolving original images with a Gaussian function of FWHM=3 pixels to make the area of objects more extended. If the original image (Figure ~\ref{fig:sub_SKY}-a) is directly used for Sextractor to detect objects, it may be hard to derive a good object-masked image because the wings of bright objects can not be completely masked \citep{2015AJ....149..199D}. Secondly, we got the object-masked image by subtracting all of the detected objects according to their masked areas from original image. Thirdly, we used method provided by \cite{2018ApJS..235...18L} to get a reliable large-scale structures of the background by using this object-masked image. We applied a median filter of $70 \times 70$ $pixel^{2}$ to convolve the object-masked image in order to reduce the random noise and to fill in the mask regions by surrounding sky region.  Unfortunately, due to the fact that our objects are too large, the backgrounds in the center of sources are not filled (Figure ~\ref{fig:sub_SKY}-b). We adopted two methods to deal with this problem.

We found that the background of our images had similar pattern in the same filter. So we conducted image combination with a 3 sigma clip to get an average background for each filter, which removed most signatures for the regions masked incompletely, such as the region unfilled as well as the wings of the bright stars. (Figure ~\ref{fig:sub_SKY}-c).  The average-background was then scaled and subtracted from the original image. Figure ~\ref{fig:sub_SKY}-d shows an example of the sky-subtracted image.

There were still some images with strange pattern that we can not get the background by this method and we made the background as did by \cite{1999AJ....117.2757Z, 2002AJ....123.1364W} and \cite{2015AJ....149..199D}. We performed a least-squares polynomial fit of low order to the sky pixels of each row and column  and then averaged the line-fitted and column-fitted images to get the averaged background. At last this averaged image (Figure ~\ref{fig:sub_SKY}-f) was smoothed with a Gaussian function of FWHM=31 pixels (Figure ~\ref{fig:sub_SKY}-g) and used as sky background. Figure ~\ref{fig:sub_SKY}-h shows the sky-subtracted example for the second method.

\begin{figure*}[!htb]
	\begin{center}
		\includegraphics[angle=0,scale=0.35,keepaspectratio=flase]{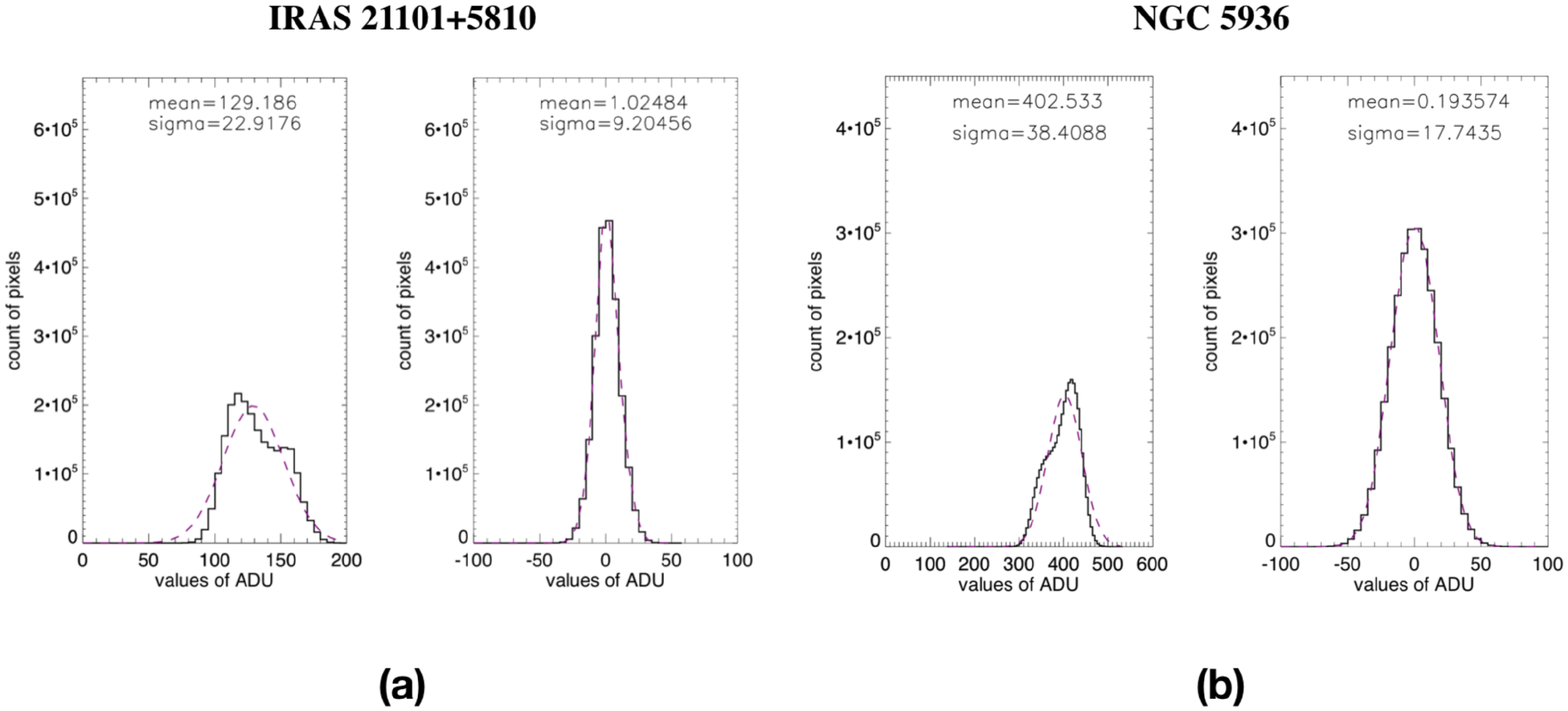}
		\caption{An example of counts distribution before and after sky subtraction. The left panels of figure (a) and (b) represent the counts distribution for all unmasked pixels in the original image of  RAS 21101+5810 in $H\alpha$-6 (method-1) and NGC 5936 in $H\alpha$-3 (method-2), respectively. The right panel of figure (a) and (b) represent the counts distribution for all unmasked pixels in the sky-subtracted image. The upper of each panel gives the mean value and standard deviation of distribution. \label{fig:sub_SKY3}}
	\end{center}
\end{figure*}

The sky background derived from these two methods both show the vignetting and non-uniformity distribution. Figure ~\ref{fig:sub_SKY3} (a) (b)shows the fluctuation of two example images for method-1 and method-2. It is clear that the mean values of the Gaussian distribution for images with our sky background subtraction are close to zero and have much less fluctuation than those of the original images.

\begin{figure*}[!htb]
	\begin{center}
		\includegraphics[angle=0,scale=0.3,keepaspectratio=flase]{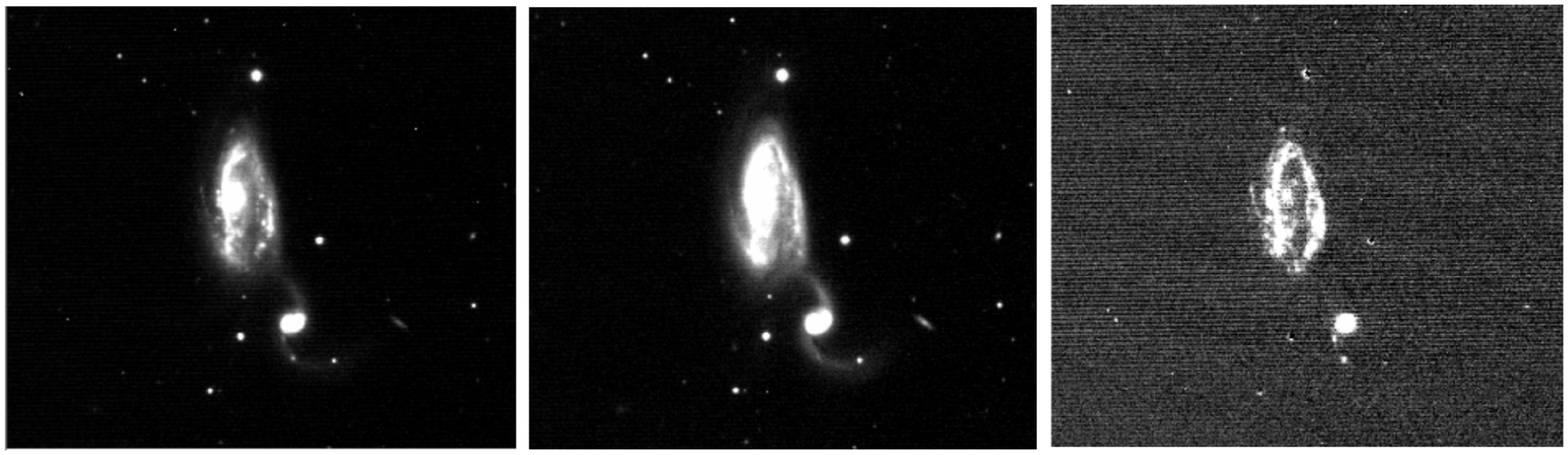}
		\caption{An example of continuum-subtracted. The $H\alpha$ , R-band and continuum-subtracted $H\alpha$ images of $LIRGs$ $NGC5394/5$ are shown in this figure from left to right. \label{fig:sub_Ha} }
	\end{center}
\end{figure*}

Then the $H\alpha$ image were scaled relatively to the continuum R-band images using field stars, and the continuum R-band images were subtracted from the scaled $H\alpha$ image to yield continuum-free images. In this process, we assume the absence of feature lines on their continua of the field stars.  The scaling factors are defined by the ratio between counts of field stars in the wide R-band and narrow $H\alpha$ band. We adopt the median value first and then adjust the value around until the residual fluxes of foreground stars reached the minimum.  Figure ~\ref{fig:sub_Ha} shows the $H\alpha$ band, R-band and continuum-subtracted $H\alpha$ images of $NGC5394/5$ as an example. 

\subsection{Photometry} 

The continuum-subtracted $H\alpha$ images were flux calibrated using photometry from \emph{Panoramic Survey Telescope and Rapid Response System (Pan-STARRS)}. The  \emph{Pan-STARRS} survey is designed for collecting wide-field astronomical imaging and operated by Institute for Astronomy at the University of Hawaii. This survey used a 1.8-m telescope with a 1.4 Gigapixel camera to image the sky in five broadband filters \emph{(g, r, i, z, y)}. The systematic errors in Pan-STARRS photometric system is about 0.02 mag \citep{2012ApJ...750...99T}. In our work, the Pan-STARRS's PSF magnitude of g-band ($m_{g}$) and r-band ($m_{r}$) were used to get the Johnson/Cousins R-band magnitude ($m_{R}$) by the formula given by \cite{2012ApJ...750...99T}:

\begin{center}
$m_{R}-m_{r}=-0.138-0.131(m_{g}-m_{r})\pm0.015$
\end{center}

\noindent Then the $m_{R}$ is transformed to flux density with following equation \citep{1983ApJ...266..713O,1994AJ....108.1476F}:

\begin{center}
$m_{AB}=m_{R}+0.055,$

$m_{AB}=-2.5log_{10}(\frac{f_{v}}{3631JY}).$
\end{center}

Then by comparing the field star in our observation and Pan-STARRS, we derived the flux calibration in this observation.

\begin{figure*}[!htb]
	\begin{center}
		\includegraphics[angle=0,scale=0.3,keepaspectratio=flase]{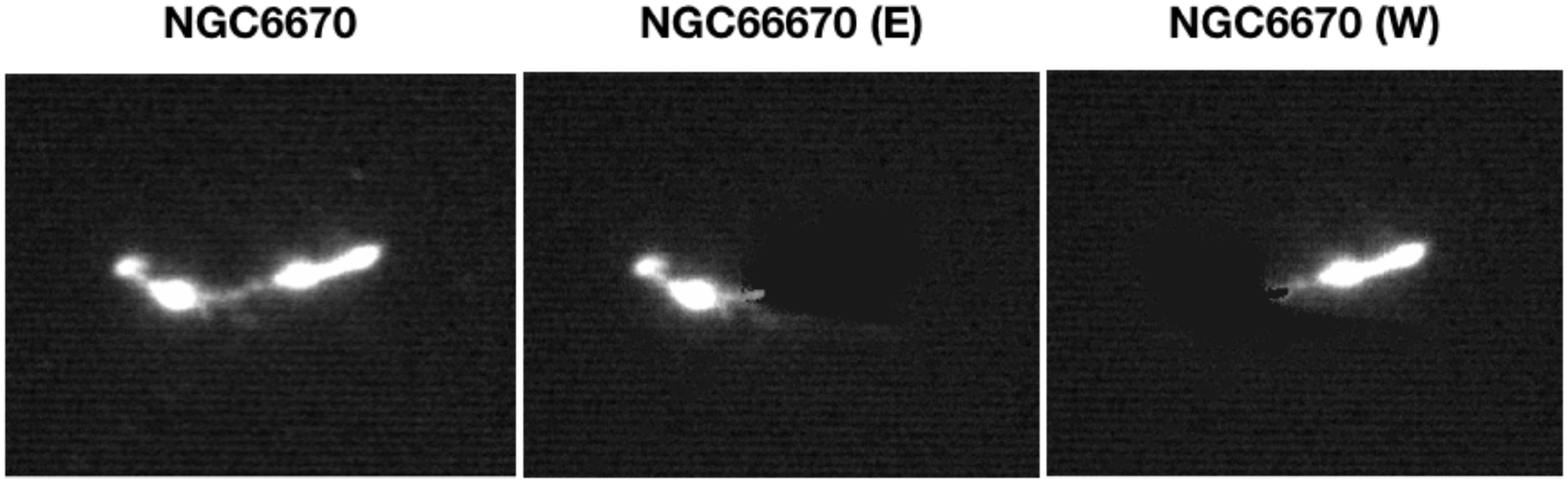}
		\caption{An example of interaction galaxies separation.  The interacting galaxies with region overlapped are separated by $Sextractor$ (NGC 6670 in R-band). \label{fig:sextractor} }
	\end{center}
\end{figure*}

Before the photometry, field stars must be masked in R-band images.  The $Sextractor$ was used to find stars across the image and replaced them with the median of background value. The counterparts of galaxy pair were masked in a similar way. When we measure one object, the other is masked, as an example be seen in Figure ~\ref{fig:sextractor}.  The continuum-subtracted $H\alpha$ images also require masking in the case of galaxy pair or the residuals from star were not subtracted clearly. The $L_{IR}$ was also assigned into two counterpart of galaxies pair according to their $H\alpha$ fluxes ratio. Although the $L_{IR}$ of some separated galaxies does not meet the requirement of LIRGs ($10^{11} L_{\sun}$), we still include these sources in our sample. 

Then, we performed the photometry with \emph{IRAF} ellipse package.  Firstly, we fitted ellipse isophotes to R-band images. The center of the galaxies were determined by the contour map of R-band images. Many objects in our sample show the features of bars, rings, and interaction disturbance. For this reason, the ellipse isophotes  were derived by allowing the position angle ($PA$), ellipticity ($e=\frac{a-b}{a}$) and galaxy center to vary along the radius during the fitting process. Starting values of ellipticity and position angle were determined by eye from the contour map of the galaxies in R-band. We derived a set of concentric elliptical isophotes which are extend from nuclear region to outskirt of  galaxies. When the variation of enclosed flux was close to zero among at last five isophotes, we used this radius as the boundary of the galaxy ($radii$ $R$).  We also get the half-light radius ($R_{e}$) at which the enclosed R-band fluxes reach the half of total. 

\begin{figure*}[!htb]
	\begin{center}
		\includegraphics[angle=0,scale=0.6,keepaspectratio=flase]{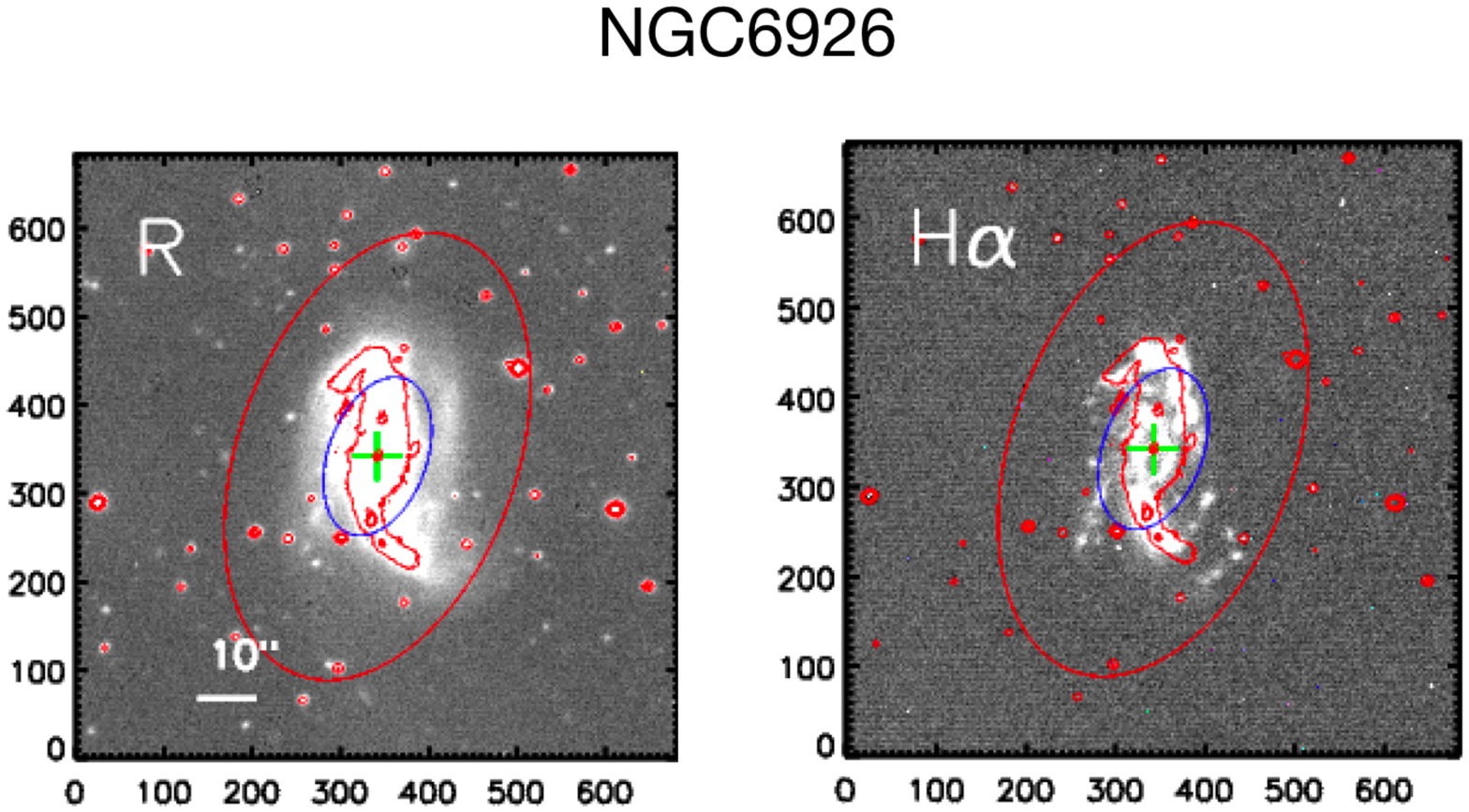}
		\caption{An example of  galaxy boundary radius. The left panel is R-band image and the right panel is continuum-subtracted $H\alpha$ image. The red ellipse represents the boundary radius ($radii$ $R$), the blue ellipse represent the $R_{e}$ radius, the green cross represents the galaxy center and the red contours  are for R-band . The solid line on the R-band images represent $10^{\prime \prime}$. \label{fig:Re} }
	\end{center}
\end{figure*}

The photometry of $H\alpha$ was measured using the ellipse isophotes obtained from R-band images.  Figure ~\ref{fig:Re} shows the two ellipses for $NGC$ $6926$, which enclosed the total flux (red one) and half flux (blue one).

The R-band and continuum-subtracted $H\alpha$ images are presented for each object in $appendix$ $A$ (Figure ~\ref{fig:R_Ha}).

\subsection{$H\alpha$ Flux Correction} 

The corrections for $H\alpha$ flux include (1) $H\alpha$ filter transmission; (2) $[NII]$ emission; (3) both Galactic and internal galaxies extinctions.

 Because some of the redshifted $H\alpha$ lines may locate at lower transmission part of filter band, we applied transmission correction to the objects as suggested by \cite{2018ApJS..235...18L}. A normalized transmission $T(H\alpha)$ was used for the correction:

\begin{center}
	$T(H\alpha)=\frac{T^{'}(H\alpha)}{\int_{\lambda1}^{\lambda2} T^{'}(\lambda)d \lambda/FWHM}$,
\end{center}

\noindent where $T'(\lambda)$ presents the transmission curve,  $T'(H_\alpha)$ is the directly transmission of redshifted $H\alpha$ emission, the $FWHM$ represent the width of $H\alpha$ filter curve at half of it's peak value, $\lambda1$ and $\lambda2$ represent the begin and end wavelength of transmission curve, respectively.  The transmission-corrected $H\alpha$ flux is obtained by dividing the $T(H\alpha)$.  In addition, the flux in R-band also contains the $H\alpha$ emission,  which will result in underestimate for $H\alpha$ flux in the process of continuum subtraction.  Such loss ($4\%$) was estimated by \cite{2018ApJS..235...18L} and corrected for our objects.

\cite{1998A&AS..132..181W} performed a spectroscopic observation of 73 LIRGs and provided ratio of $[NII]/H\alpha$ of this sample. We took the mean value (0.55) of these LIRGs and then used it to correct the $[NII]$ emission in our sample.   

The Galactic extinction was corrected by using the \cite{1998ApJ...500..525S} map and the extinction curve from \cite{1999PASP..111...63F}. There are several methods for estimating the intrinsic extinction of $H\alpha$. \cite{1996AJ....112.1903Y} derived this correction from $[SIII]/H\alpha$ ratio.  \cite{2016ApJ...822...45T} assumed a correction of 1 magnitude based on $L_{H\alpha} $-$ SFR$ relation.  The internal extinction in work of \cite{2012A&A...545A..16G} were performed by using Balmer decrement. As there were no spectral emission lines which we can used for internal extinction correction, we adopted the mid-IR luminosity to estimate the intrinsic extinction of $H\alpha$ flux. \cite{2008ApJ...686..155Z} presented a correlation between spitzer 24 $\mu m$ mid-infrared and extinction-corrected $H\alpha$ luminosities for star-forming galaxies. By combing the $H\alpha$ emission line and 24 $\mu m$ measurements for nearby galaxies, \cite{2009ApJ...703.1672K} got the similar relation and derived the formula. We chose their formula as:

\begin{center}
	$ A(H\alpha)(mag)=2.5log [1+\frac{0.020L(24)}{L(H\alpha)_{obs}}]$,
\end{center}

\noindent where $L(24)$ is spitzer mid-infrared luminosity (MIR) at 24 $\mu m$ (here we adopted as IRAS 25 $\mu m$ instead) and the $L(H\alpha)_{obs}$ is observed $H\alpha$ luminosity without internal extinction correction.  

The main errors of $H\alpha$ fluxes include photometry and continuum-subtraction. The photometric errors due to the $H\alpha$ photon counting noise and background noise are typically smaller than 4$\%$.  The scaling factor of continuum-subtraction is the dominant source of uncertainty. Even small uncertainties in scaling factor can result in large uncertainties in $H\alpha$ flux with relatively weak $H\alpha$ emission. We produced continuum-subtracted $H\alpha$ images with a range of scaling factor. And then the accuracy of scaling factor was estimated by the value at which the continuum-subtracted $H\alpha$ image are clearly oversubtracted and undersubtracted. The typical errors in continuum subtraction is around 25$\%$ and in few exceptional case, this error reaches 70$\%$. By the way, the errors of the internal extinction correction is mainly composed of two parts. One is the uncertain of extinction correction formula (typically 15$\%$) and another is the errors of IRAS $25\mu m$ fluxes (typically 5$\%$).

\section{$H\alpha$ Imaging Results} 

In this section, we present the primary results of $H\alpha$ imaging observation. The Table~\ref{table:Ha}, together with Figure ~\ref{fig:R_Ha}, constitutes the main results of our observation.

\subsection{$H\alpha$ catalog}

The $H\alpha$ photometry result of 158 galaxies are listed in Table~\ref{table:Ha}. Both $H\alpha$ Luminosity before and after internal extinction correction are given, as well as the ratio between $H\alpha$ flux enclosed in $R_{e}$ and that of total galaxy.

Column (1): Source name;

Column (2): Type.  ---The morphology and interactions type of (U-)LIRGs (the detail description will be given in next section);

Column (3):  The $L(H\alpha)_{obs}$ .  ---The observed $H\alpha$ luminosity after corrected for transmission, [NII] emission and Galactic extinction in the units of $erg$ $s^{-1}$;

Column(4):  The $L(H\alpha)$.  --- The $H\alpha$ luminosity after corrected for internal extinction in the units of $erg$ $s^{-1}$;

Column(5):  $Frac(H\alpha)$.  ---The ratio between $H\alpha$ flux enclosed inside $R_{e}$ and the total;

Column(6):  $PA$.  ---The adopted position angles at galaxies boundary ($radii$ $R$).

Column(7):  $e$.  ---The adopted ellipticity at galaxies boundary ($radii$ $R$).

\begin{figure}[!ht]
	\begin{center}
		\includegraphics[angle=0,scale=0.5,keepaspectratio=ture]{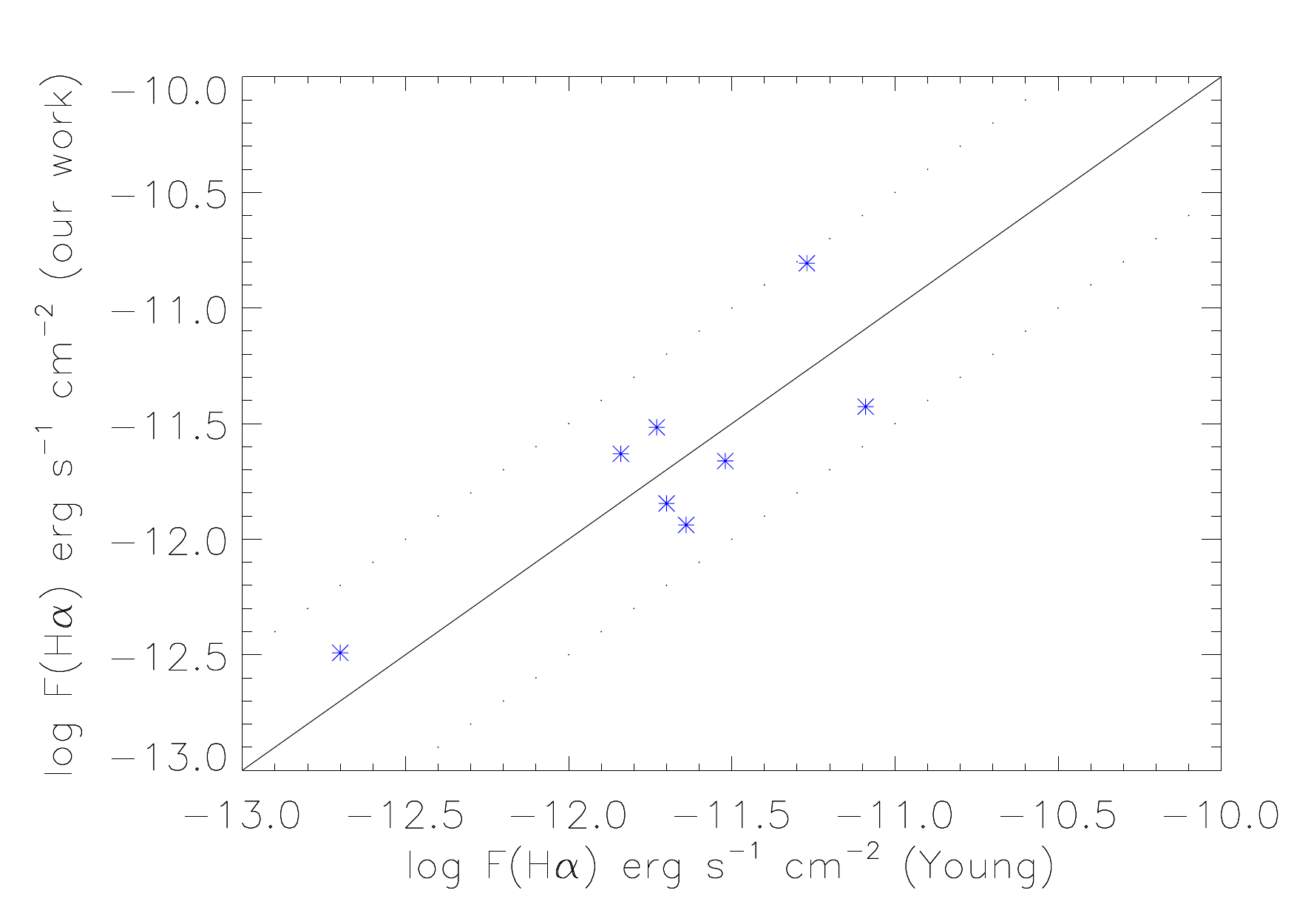}
		\caption{ Comparison of $H\alpha$ fluxes for galaxies in our work with \cite{1996AJ....112.1903Y}. }
		\label{fig:compare} 
	\end{center}
\end{figure}

Figure ~\ref{fig:compare} shows a comparison of $H\alpha$ emission flux measured by us with those of \cite{1996AJ....112.1903Y}. The objects of  \cite{1996AJ....112.1903Y} were measured without the correction of $[NII]$ emission, internal galaxies extinction and Galactic extinction. In the comparison, we do the same steps as their work and give the comparison result.  All objects show a good agreement around 0.24 dex.

\subsection{$H\alpha$ images}
Figure ~\ref{fig:R_Ha}, in $appendix$ $A$, presented the R-band and continuum-subtracted $H\alpha$ images for all 158 objects. These images were reduced by using standard IRAF task. The WCS parameters in the FITS header were added using Astrometry.net. We calibrate the images by adding the flux calibration scale value to the imaging header as ``scale". The counts value can then scaled to flux ($erg\, cm^{-2}\, s^{-1}$) by multiplying this value. This ``scale" value also include the correction for the $H\alpha$ filter transmission and 4\% underestimate for $H\alpha$ flux. We didn't make $[NII]$ emission, Galactic and internal galaxies extinction corrections for the ``scale" value. These images are listed in order of object name and the solid line on the R-band images represent $10^{\prime\prime}$.  All these images in FITS format can be download via the ApJS website.

\section{Morphology Analysis}

\subsection{Morphology Classification}

We divided the sample into several morphological classes in order to understand their role in the evolution of galaxies. We made our own morphological classification based on R-band as follows: 

S (Spiral) ---  spiral galaxies with symmetrical disk and shows no signs of tidal interaction;


$PM$ (Pre-Merger) --- two galaxies can be separated with asymmetrical disks or tidal tails, which could be the phase before merging;

M (Merger) --- galaxies contain two nuclei with tidal tail or the galaxies are disturbed severely which is associated with most violent dynamical events;

LM (Later Stage of Merger) --- single nucleus with short faint tidal tail which may in the late stage of merger;

E (Elliptical Galaxy) --- elliptical galaxies with an approximate ellipsoidal shape and without tidal interaction, which could be in final phase of merger;

UN (Uknown) --- objects can not  classified by their  morphology clearly.

\begin{figure*}[!h]
	\begin{center}
		\includegraphics[angle=0,scale=0.6,keepaspectratio=flase]{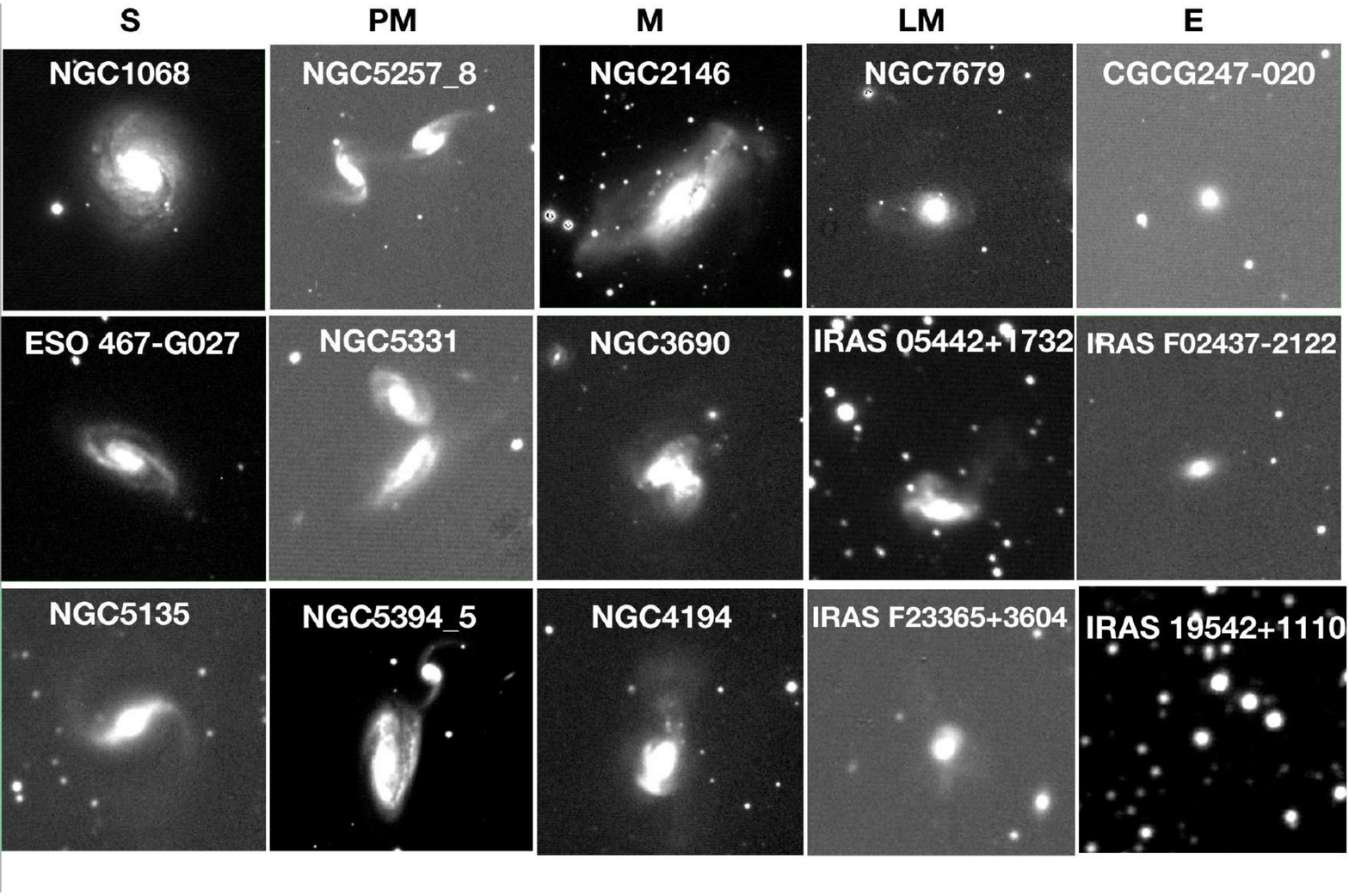}
		\caption{ Example galaxies which are grouped into five morphological types. S: spiral galaxies; PM: galaxy in the predecessors of merger stage; M: galaxies in most violent merge stage; LM: galaxies in late merge stage; E: elliptical galaxies. In all case north is up and east is left.  \label{fig:classify} }
	\end{center}
\end{figure*}

\begin{figure}[!h]
	\begin{center}
		\includegraphics[angle=0,scale=0.3,keepaspectratio=flase]{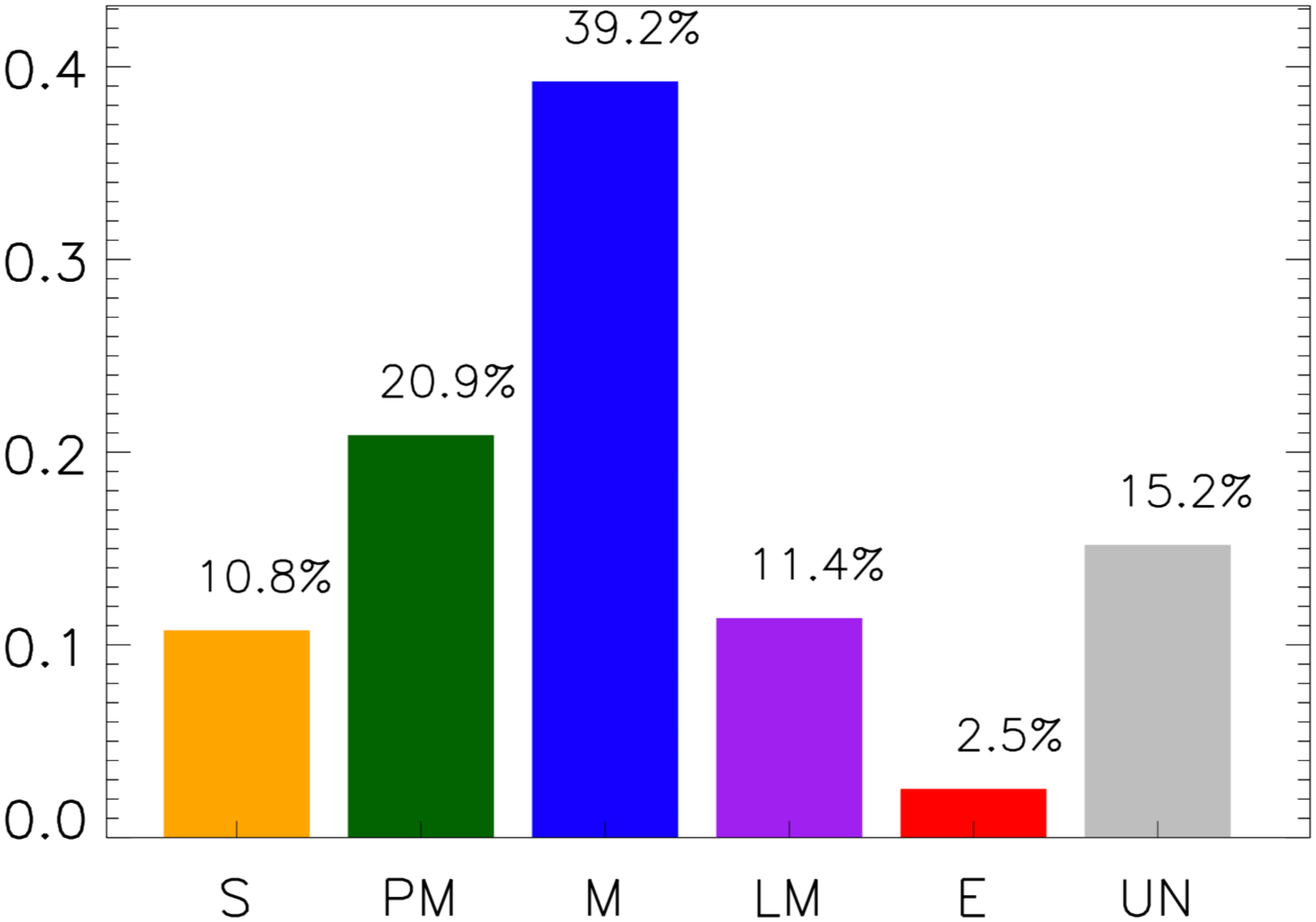}
		\caption{ Distribution of morphology type: S (orange); PM (green), M (blue), LM (purple); E (red) and UN (gray) \label{fig:type} }
	\end{center}
\end{figure}

Examples of different morphological classes are given in Figure ~\ref{fig:classify}. The classification was done independently by different people and a consistent classification was adopted after deliberated discussions. In our classification, the $E$-$type$ occupies the smallest fraction ($2.5\%$) among morphological types. The $M$-$type$ occupy the largest fraction ($39.2\%$).  (U-)LIRGs in $S$-$type$, $PM$-$type$ and $LM$-$type$ occupy the percentage of 10.8, 20.9 and 11.4, respectively. The $UN$ occupies 15.2\%. The Figure ~\ref{fig:type} shows the distribution of morphological class in histograms.

\subsection{Infrared Luminosity}

\begin{figure}[!h]
	\begin{center}
		\includegraphics[angle=0,scale=0.5,keepaspectratio=flase]{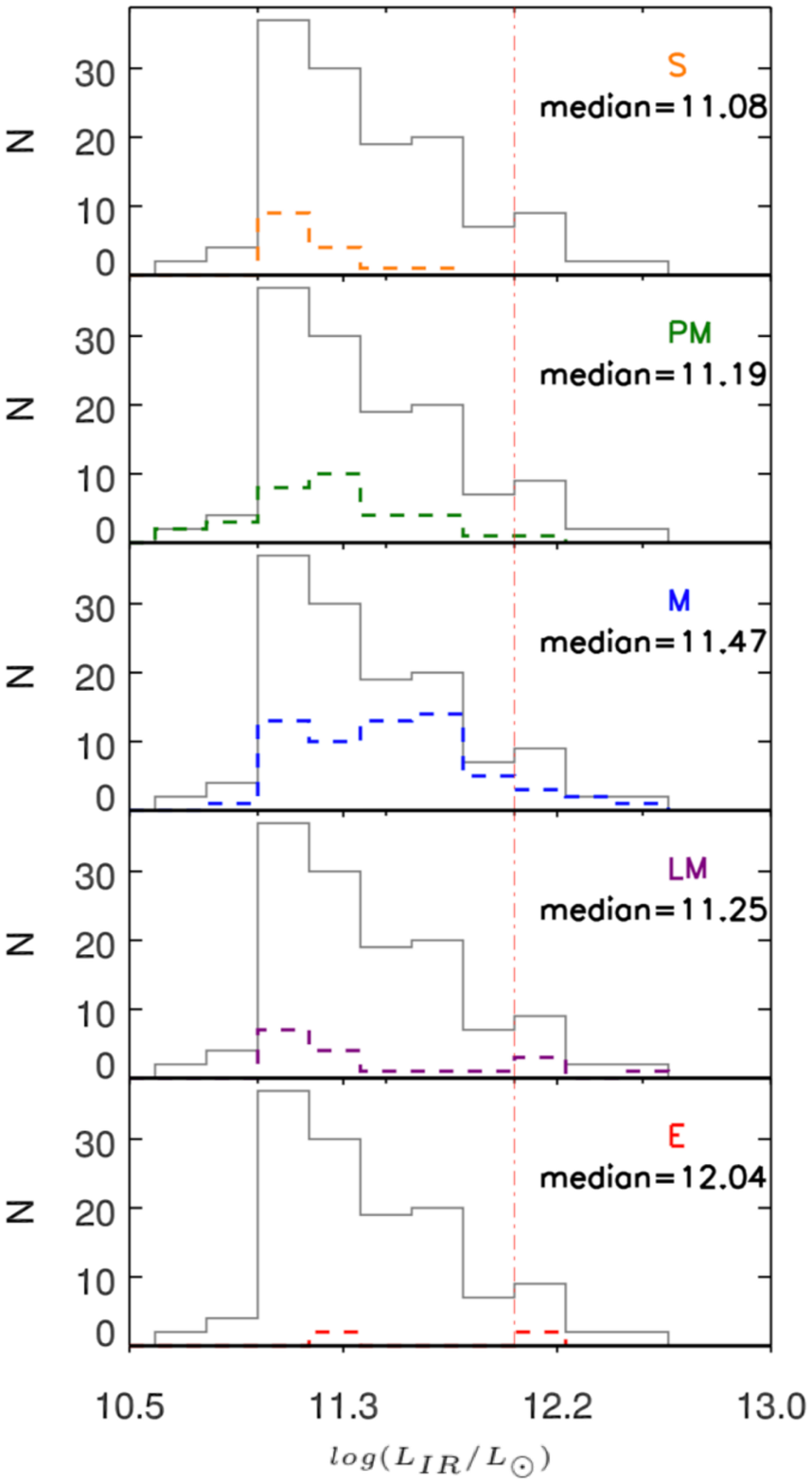}
		\caption{ The $L_{IR}$ distributions of different morphology type. The  boundaries of LIRGs and ULIRGs are showed with vertical line (red dot-dashed line).  Colors are same as those in previous figure and the black lines show the distribution for the whole sample. The median vale of $L_{IR}$ for each type is also shown in the up right corner of the each panel. \label{fig:IR_Ha} }
	\end{center}
\end{figure}

The Figure ~\ref{fig:IR_Ha} shows the distributions of $L_{IR}$. The black lines show the distribution for the whole sample.  The distributions for other morphology types are also given in this figure ($S$: orange,  $PM$: green, $M$: blue, $LM$: purple and $E$: red).  
As can be seen, the $S$-$type$ appears to be skewed toward smaller value and none of them have $L_{IR}$ larger than $10^{11.65} L_{\sun}$, which is consistent with previous works \citep{2006ApJ...649..722W, 2015RAA....15.1424L, 2016ApJ...825..128L}. 

The median $L_{IR}$ in $PM$-$type$, $M$-$type$, $LM$-$type$ and $E$-$type$ is showed in each panel of Figure ~\ref{fig:IR_Ha}. And as the merging process advanced the objects have a tendency to have relatively extended tail toward larger $L_{IR}$. \cite{2004ph.D} showed the same result in their study of 56 LIRGs that the separation between merging galaxies decrease as IR luminosity increases.

\subsection{$H\alpha$ Luminosity and Concentration}

\begin{figure}[!h]
	\begin{center}
		\includegraphics[angle=0,scale=0.5,keepaspectratio=flase]{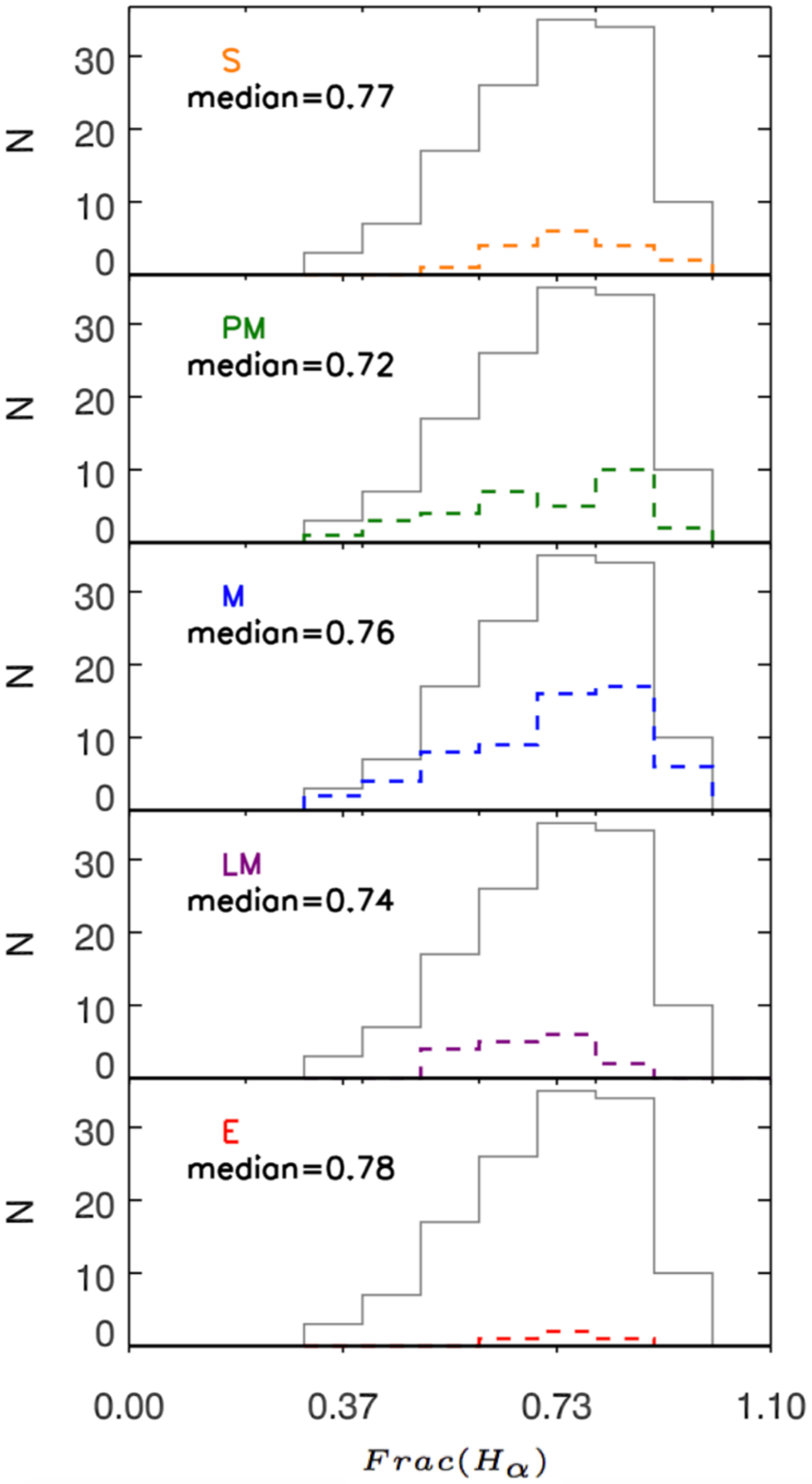}
		\caption{ The distribution of $Frac(H\alpha)$. Colors are same as those in previous figure. The median vale of $Frac(H\alpha)$ for each type is also shown in the up left corner of the each panel.\label{fig:Ha_frac} }
	\end{center}
\end{figure}

In Figure ~\ref{fig:Ha_frac}, the histogram of $Frac(H\alpha)$ is showed for each morphological type. It is clear that for most (U-)LIRGs, the star-forming is dominated by the central region with $Frac(H\alpha) > 0.5$. The $S$-$type$ have a moderate concentration among all types with a median value of 0.77.

The $Frac(H\alpha)$ is also expected to be higher following the advancing of merging process \citep{1999AJ....117.2632B,2004AJ....127..736H}. The $Frac(H\alpha)$ of $PM$-$type$ is the minimum, and increase along the merging sequence from M-, LM- to E-type. The median value of $Frac(H\alpha)$ is showed in each panel of Figure ~\ref{fig:Ha_frac}.

\subsection{$H\alpha$ Profile}

\begin{figure}[!h]
	\begin{center}
		\includegraphics[angle=0,scale=0.5,keepaspectratio=flase]{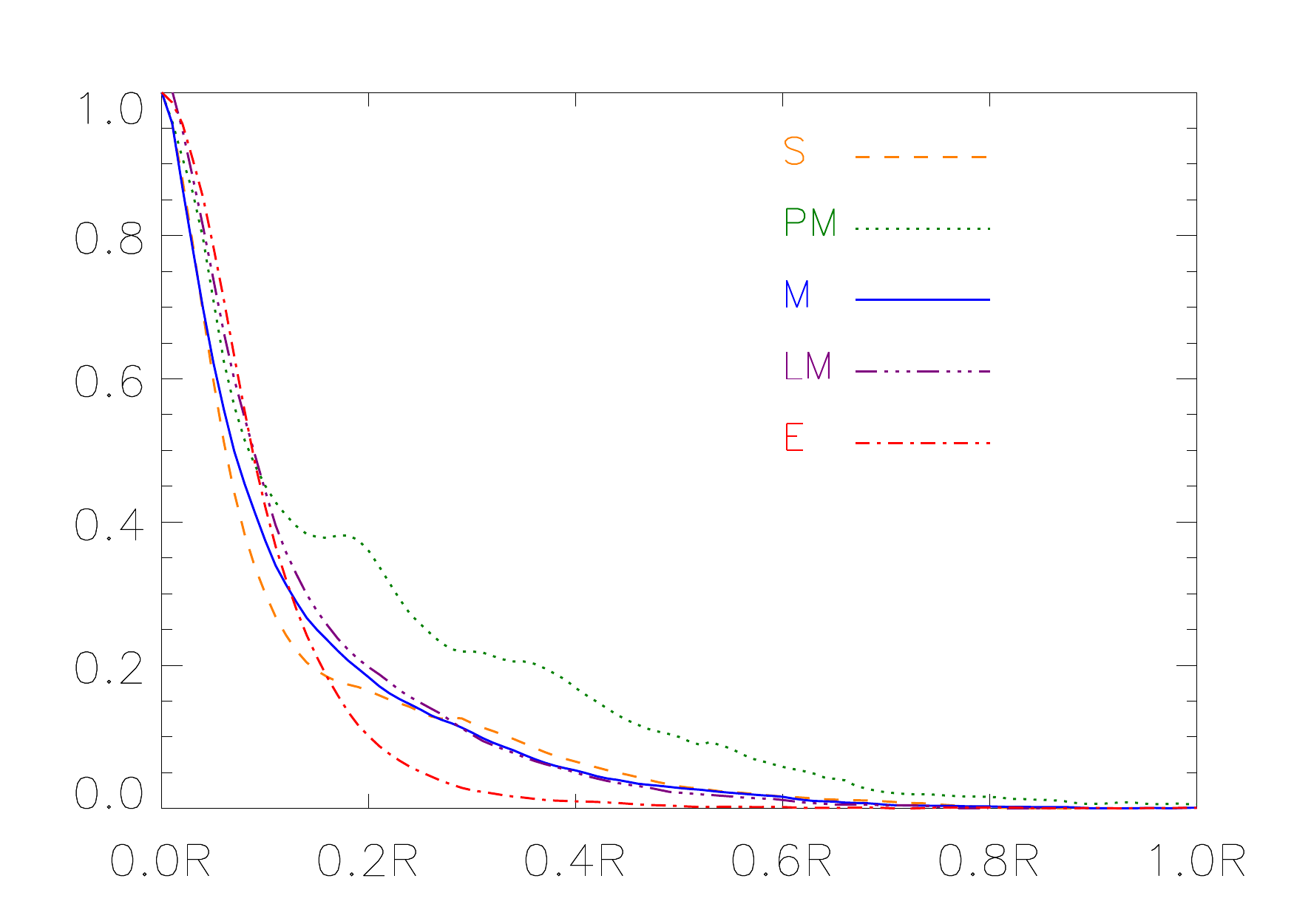}
		\caption{ The isophotal shape of different morphology type. The orange dashed line represents the $S$ -$type$; the green dotted line represents the $PM$-$type$; the blue solid line represents the $M$-$type$; the purple dotted-dotted-dashed line represents the $LM$-$type$ and the red dotted-dashed line represents the $E$-$type$. The horizontal coordinate is scaled with galaxies boundaries (R). \label{fig:isophotal} }
	\end{center}
\end{figure}

Figure ~\ref{fig:isophotal} shows the profile for various morphology types. The 10 edge-on galaxies (S:2; I:6, M:1, LM:1) are not involved.  In the direction of intensity, we normalize the profile with center intensity. In the direction along the galaxy's radii, we normalize the profile with galaxy boundary $radii$ $R$. Then we combine the profiles according to their morphology types.

The $PM$-$type$ are characterized by exceptionally extend profile. The $S$-$type$ also shows some extensions in the outer region.  The $PM$-$type$ galaxies may be dynamically young system  which are predecessors for advanced merger stage.  On the other hand, the $E$-$type$ galaxies which are relaxed from interaction without sign of interaction are the most compact one. The $M$-$type$ and $LM$-$type$ are similar and have the  intermediate profile among others. 

By using the mid-infrared emission of LIRGs which can show the structures for different merge stage, \cite{1999ApJ...511L..17H} found out that the peak-to-total flux ratios of LIRGs increase as projected separation of interacting galaxies become smaller. The profile as well as $Frac(H\alpha)$ in our study are consistent with previous studies that (U-)LIRGs tend to have the more concentrated star-formation distribution as the merging process advance \citep{1999AJ....117.2632B,2004AJ....127..736H}.

\subsection{Infrared Color}

The infrared colors have been interpreted by various models \citep{1986ApJ...311L..33H,1987ApJ...316..145S}. A cool component temperature ($20K$) is used to represent the emission from dust in infrared cirrus heated by older stellar population and peaks at $\lambda \gtrsim 100 - 200 \mu m $.  A warmer component temperature ($30 \sim 60 K$) represents the starburst in galaxies and peaks near $60 \mu m$. And a even warmer component peaking around $25 \mu m$, represents the dust emission heated by AGN.

\begin{figure*}[!h]
	\begin{center}
		\includegraphics[angle=0,scale=0.6,keepaspectratio=flase]{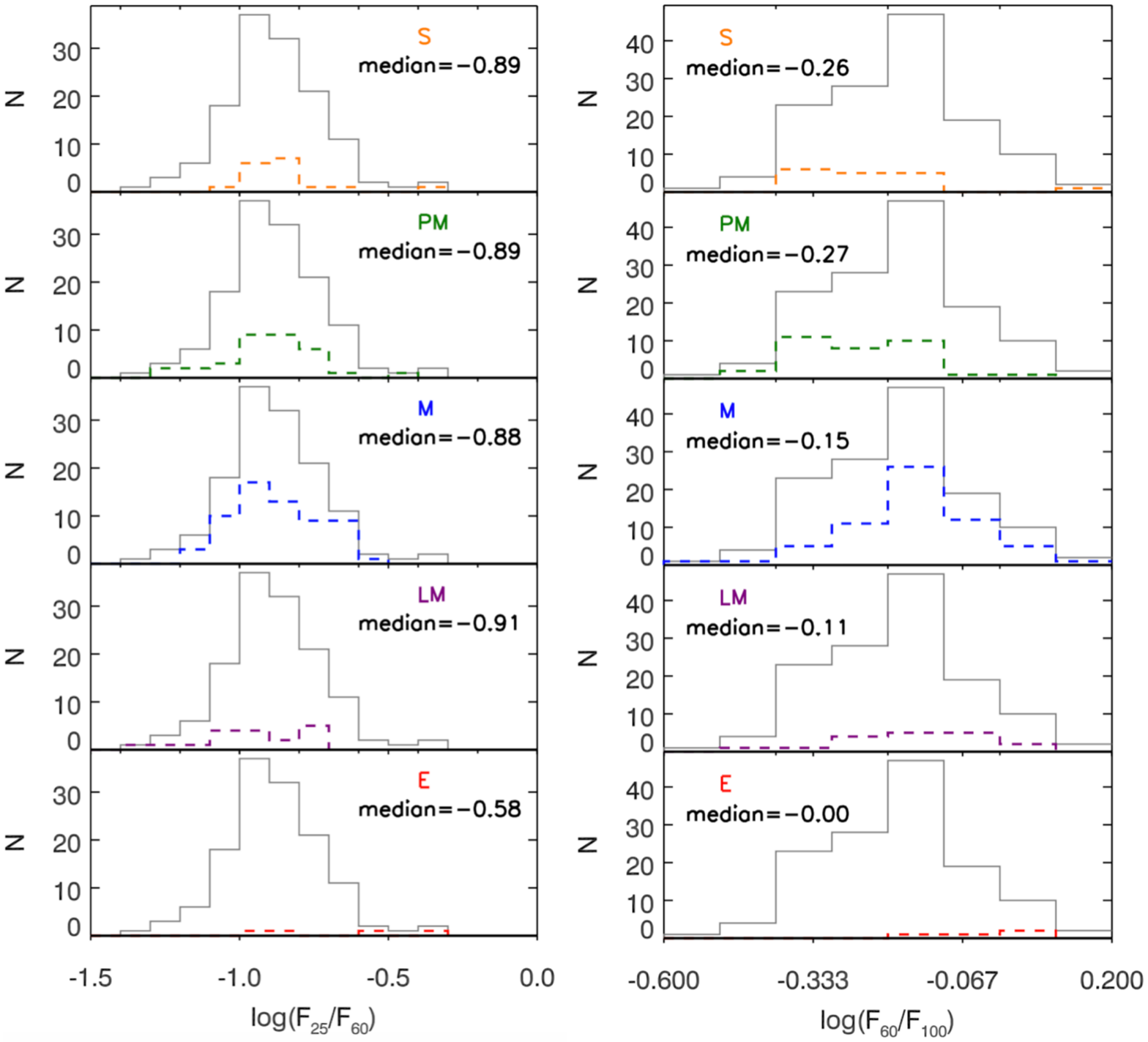}
		\caption{ The IR-color of different morphology type. Colors are same as those in previous figure. The median vale of  IR-color for each type is also shown in each panel.\label{fig:color} }
	\end{center}
\end{figure*}

The distributions of IR color ($log(f_{25}/f_{60})$ and $log(f_{60}/f_{100})$ ) is showed in Figure ~\ref{fig:color}. The $E$-$type$ have a warmer $f_{25} / f_{60}$ which may indicates they host an AGN. The $NGC1068$, a spiral galaxy, which also has warmer $f_{25} / f_{60}$ (-0.35) is a well-known Seyfert2 galaxy. The rest objects don't have clearly tendency in $f_{25} / f_{60}$. In addition, (U-)LIRGs  tend to have warmer  $log(f_{60}/f_{100})$ as the merging process advanced.
The median values of both  $log(f_{60}/f_{100})$ and $log(f_{25}/f_{60})$ are also showed in each panel of Figure ~\ref{fig:color}.

\section{Discussion} 

\subsection{Importance of a Complete $H\alpha$ Imaging Survey }
The GOALS sample has become a ``reference sample" for studying the properties of (U-)LIRGs in the local universe. Extensive multi-wavelength (radio to X-ray) imaging and spectroscopic data have been obtained for different subsamples of GOALS. Although the Balmer $H\alpha$ emission line is a good indicator for star formation rate and there are some $H\alpha$ imaging observation containing sources belong to GOALS sample, until now, there is no complete $H\alpha$ observation data for this ``reference sample". 
	
 In this work, we performed an $H\alpha$ survey for a complete GOALS subsample with $Dec. \ge -30^{\circ}$. After continuum-subtracted, we obtain 148 pure $H\alpha$ emission images, which can provide the star-formation distribution for this GOALS subsample.  We also provide a relatively complete $H\alpha$ photometric data for GOALS subsample for the first time. In sum, this survey provides the imaging and photometry component which is a useful data addition to the GOALS data archive, and is helpful in revealing the formation and evolution of (U-)LIRGs.

 \subsection{Comparison with Other Morphological Classifications}
 
 Though there are many previous works focus on morphological classifications (eg.  \cite{2011AJ....141..100H}; \cite{2013ApJ...768..102K} and \cite{2016ApJ...825..128L} ), their data are not completely covered objects in this work and the classification results cannot be directly used. Here we don't use the same classification criterion as they do. Previous studies of LIRGs morphology either relied on HST higher resolution ($\sim 0.1^{\prime\prime}$) imaging (\cite{2016ApJ...825..128L}) or mainly focus on merger stages (\cite{2013ApJ...768..102K} and \cite{2016ApJ...825..128L} ). Such as  there are no distinction between spirals and ellipticals in the classification of single galaxies  in \cite{2016ApJ...825..128L} and \cite{2013ApJ...768..102K}'s work. Considering these factors, as well as intending to distinguish the merge stage clearly, we adopt a simple approach focusing on the classification for most important merge stages.

 To ensure the reliability of morphology classification in this work, we compare our results (J19) to that of \cite{2016ApJ...825..128L} (L16) and \cite{2013ApJ...768..102K} (K13) in Figure ~\ref{fig:Mor_com1} and Figure ~\ref{fig:Mor_Com2} with method provided by \cite{2016ApJ...825..128L}. It should be noted that some merge stages, like minor merge (m) in L16, have no suitable analogs in our works. Moreover, because of different criteria for classification, as well as difference in division of stages, there will be cases where one merge stages corresponds to multiple adjacent stages. For example, the merge stage M in our work corresponds to M3 and M4 in L16, and merge stage M4 in L16 correspond M and LM in our work.

\begin{figure}[!h]
 	\begin{center}
 		\includegraphics[angle=0,scale=0.59,keepaspectratio=flase]{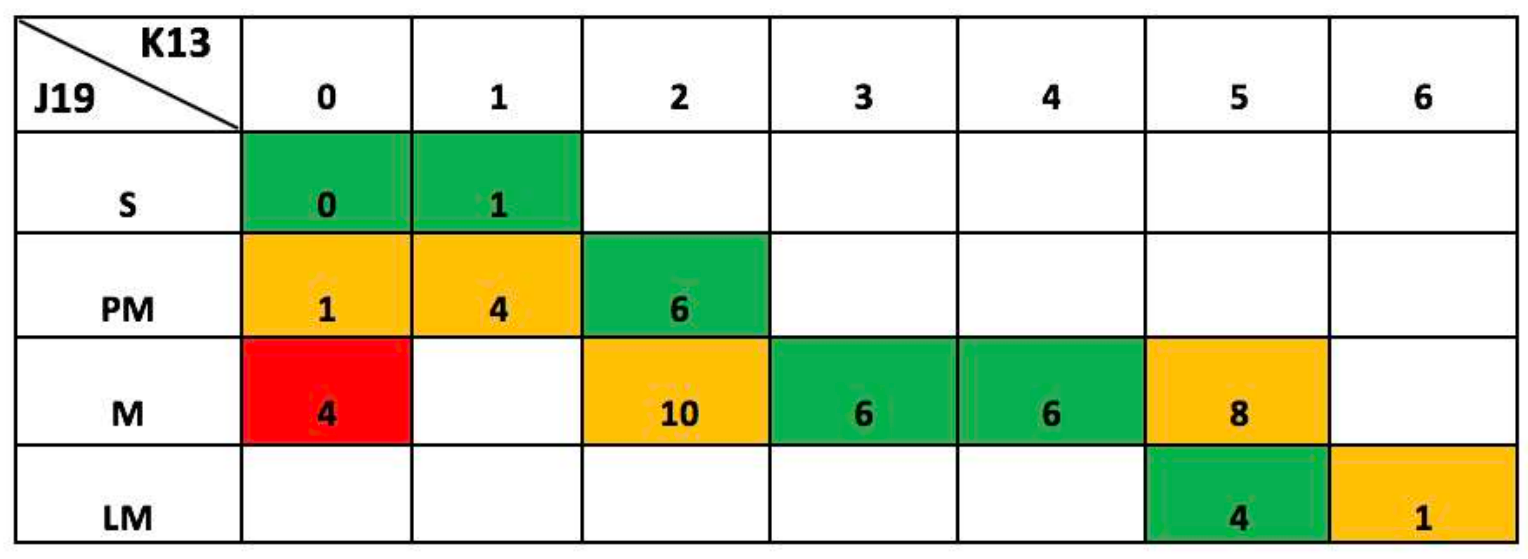}
 		\caption{ The morphology comparison between this work (J19) to \cite{2016ApJ...825..128L} (L16) \label{fig:Mor_Com1} }
 	\end{center}
 \end{figure}

\begin{figure}[!h]
 	\begin{center}
 		\includegraphics[angle=0,scale=0.55,keepaspectratio=flase]{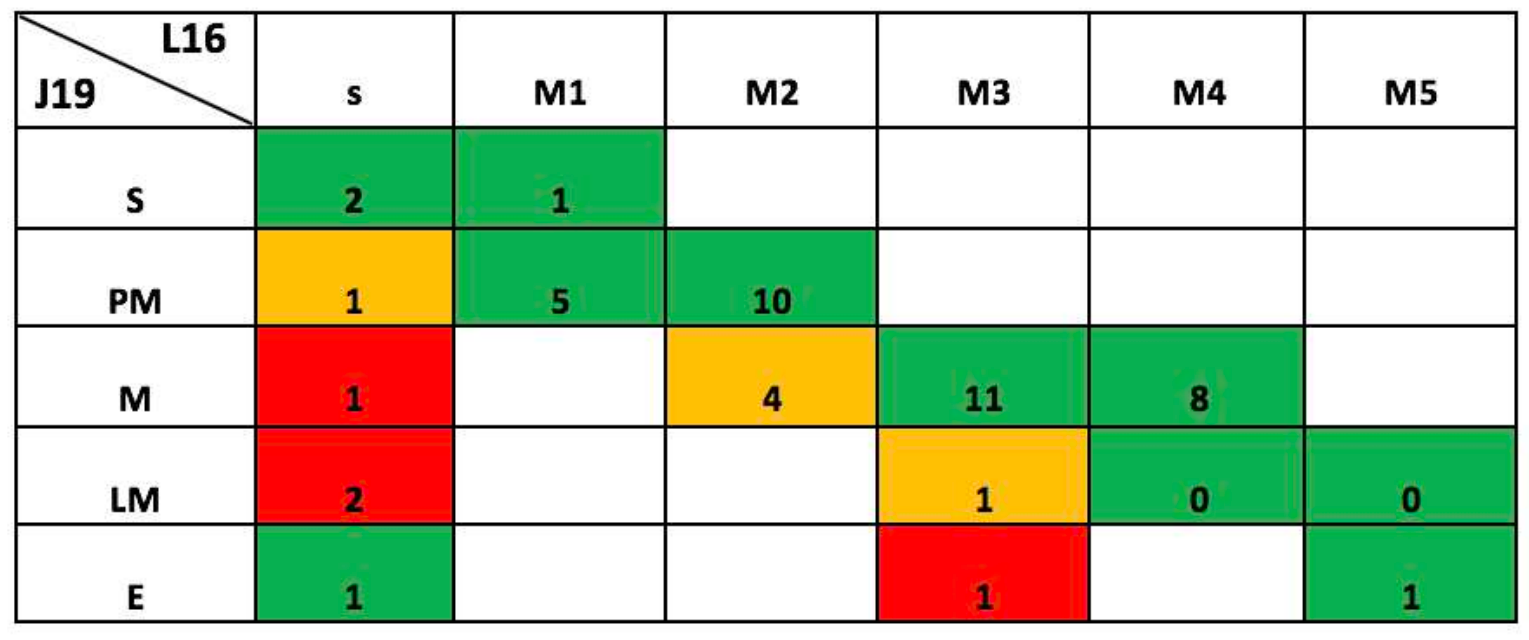}
 		\caption{ The morphology comparison between this work (J19) to \cite{2013ApJ...768..102K} (K13) \label{fig:Mor_Com2} }
 	\end{center}
 \end{figure}

 In Figure ~\ref{fig:Mor_Com1} and Figure ~\ref{fig:Mor_Com2}, we mark the cells as green when the corresponding merger stages are agree between two works. The classifications that are shifted by only one stage to earlier or later stage are marked as yellow.  Considering the morphology classification focusing on different characteristics, we treat this case as a consistent result.  When the classifications differ more than one stages, they are marked as red. There are total 63 objects are both included in our work and L16. But 10 of them don't have very certain classification nether in our work (UN), nor in L16 (ambiguities), and 4 of them are classified as minor merger (m) in L16. At last there are total 49 objects in comparison. It can be seen from Figure ~\ref{fig:Mor_Com1} that our classification agrees fairly well with those of L16 as 79.5\%  objects have very consistent classification (green) and 12.2\% objects have a slight change (yellow). And there are still 4 objects require a change more than a single stage. Overall, our classification are very consistent with that of L16. The 4 objects which require a change more than a single stage are described in the appendix B.  61 objects in our sample are also previously classified by K13 (Figure ~\ref{fig:Mor_Com2}). 7 of them don't have very certain classification in our work, and at last there are  54 objects in comparison.  The result of Figure ~\ref{fig:Mor_Com2} also shows that our classification are consistent with that of L13 (~92.6 \% are roughly the same: 46.3\%  objects have very consistent classification. 46.3\% objects have a slight change between our work and K13). The 4 objects which require a change more than a single stage are described in the appendix B. The reason why a few objects differ in our classification with that of L16  or K13 is due to different resolution of image and subjective factors. Finally, in order to maintain the consistency of our sample, we don't change our classification during the fellow analysis.

\subsection{Isolated spiral LIRGs}

In our study, the local (U-)LIRGs are dominated by merging system and the $S$-$type$ only occupies $10.8\%$ among all types.  While in \cite{2004A&A...421..847Z}'s work, $\sim 36\% $ of LIRGs with redshift between 0.4 and 1.0 are classified as normal disk galaxies. Such fraction is about three times of the fraction in our local sample. The work of \cite{2005ApJ...632L..65M} confirmed a decrease in the fraction of spiral galaxies in LIRGs from the higher redshift ($\sim$ 1) to lower redshift.

As can be seen in Figure ~\ref{fig:IR_Ha}, the $L_{IR}$ of spirals galaxies in our sample are no more than $10^{11.65}L_{\odot}$ which is consistent with previous works. \cite{2006ApJ...649..722W} also found that none of their spiral LIRGs have $L_{IR}$ higher than $10^{11.6}L_{\odot}$.  Both our and \cite{2006ApJ...649..722W} results suggest that the infrared luminosty of all the local $S$-$type$ LIRGs are much lower than the  boundary of ULIRGs ($L_{IR} = 10^{12} L_{\sun}$). And in \cite{2015RAA....15.1424L}'s work, the spiral galaxies also tend to have lower $L_{IR}$. \cite{2016ApJ...825..128L} presented an analysis of morphologies for 65 LIRGs in GOALS sample. They  found that the sources with $log(L_{IR}/L_{\sun})\ge 10^{11.5}$ are dominated by major mergers between gas-rich spirals, and all ULIRGs are late-stage merger. All these results mean that the spiral LIRGs tend to have lower $L_{IR}$ than those in merging systems in local universe. The objects in $S$-$type$ also have moderate concentration of star-forming distribution in our sources. It seems that in the local universe, the galaxies have less gas than their counterpart at intermediate redshift. And without interaction, such kind of galaxies can not enhance the extreme starburst. At intermediate redshift, the $L_{IR}$ of normal disk galaxies could be higher, because of their more gas and corresponding more extensive disk star formation \citep{2006ApJ...644..792R}. The decrease of disk gas could be one of the keys to explain the decrease in fraction of spiral galaxies in LIRGs from intermediate redshift to local universe.

\subsection{Merging sequence}

As normal spiral galaxies can not reach a higher $L_{IR}$, a merger/interacting process is needed to induce a extreme nuclear starburst. Here we explore the possible merging sequence according to our classification.

When two gas-rich spiral galaxies start their interaction, the tidal torques begin to lead a inflow of gas from outer region to central region. At this stage ($PM$-$type$ ), star-formation occurs in the both nuclear and outer region. The objects also tend to show a relatively low concentration of $H\alpha$ (Figure ~\ref{fig:Ha_frac}) and a more extended $H\alpha$ profile (Figure ~\ref{fig:isophotal}) with lower $L_{IR}$ and colder IR-color. It is consistent with evolutionary sequence described by \cite{2004AJ....127..736H}.  In their study, the objects in the early stage of interaction have a significant star formation contribution from outer region.  \cite{2001ASPC..235..281S} showed that the interaction/merging increases the colud-cloud collision which lead a transport of molecular from ISM to nuclear region and triggers the star burst in the overlap region of galaxies disks. 

As the merging process advanced ($M$-$type$ and $LM$-$type$), two nuclei of galaxies are closer. The gas continues to fall into the galaxy center and fuel the intense star-formation activity.  As a result, the star-formation activities become more active and begin to concentrate toward galaxy center. The results of previous section indicate that these objects in this stage have more concentrated star-formation region, higher $L_{IR}$ and warmer dust temperature (larger $f_{60}/f_{100}$) which  is caused by more intensive star-formation activity.  \cite{1998A&AS..132..181W} showed the same result that both infrared luminosity and $H\alpha$ equivalent width (EW) increase as galaxy-galaxy nuclear separation decreases. In the work of \cite{1992A&A...259..462L} and \cite{2004AJ....127..736H}, they both showed a warmer IR-color in this later merging stage. Many works also showed that in later merging stage, the distribution of star-formation gradually towards the center \citep{1998ApJ...500..619M,1999ApJ...511L..17H,2000ApJ...541..644X,2004AJ....127..736H}.  All these evidences indicate a intense nuclear starburst in this later merging stage.

In the last stage ($E$-$type$), the objects need to take some time to relax and toward to elliptical morphology. Such they not only shows the most concentrated distribution of star-formation region, highest $L_{IR}$ and warmest $f_{60}/f_{100}$ color but also have statistically warmer $f_{25} /f_{60}$ color. This means $E$-$type$ may have strong and concentrated star-formation activities as well as an AGN in their galaxies center \citep{1999A&A...349..735Z}. It is a stage of coexistence of both star formation and AGN. This result is consistent with classical evolution from (U-)LIRGs to QSOs.

In a word, this work shows that as the merging process advanced from PM-, M-, LM- to E-type, the objects tend to present higher $L_{IR}$, more concentrated star-formation and warmer IR-color.   All these properties support the evolutionary sequence in the (U-)LIRGs of many former works \citep{1988ApJ...325...74S,1992ARA&A..30..705B,1999AJ....117.2632B,2008ApJS..175..356H,2018ApJ...864...32J}.

\section{Summary} 

In this paper, we have presented $H\alpha$ imaging observation for a complete subsample for GOALS with $Dec. \ge -30^{\circ}$. The observation was carried out using 2.16-m telescope at the Xinglong Station of the National Astronomical Observatories, CAS, during the year from 2006 to 2009. The data present here are that so far most complete $H\alpha$ imaging survey of the GOALS sample. For many of these objects, this paper presents the first imaging data and photometry of $H\alpha$ emission.

1) There are total 148 (U-)LIRGs were observed during the $H\alpha$ imaging survey. Given there are 10 galaxies systems, our sample contain 158 galaxies at last. The subsequent data reduction mainly contains sky-background, continuum-subtracted, flux calibration, photometry and the correction of $[NII]$ emission, filter transmission, galactic extinction and internal extinction.  Finally, we obtained the $H\alpha$ images (Figure ~\ref{fig:R_Ha}) and luminosity catalog (Table~\ref{table:Ha}) for this sample.

2) We have visually classified our sample using a simplified classification which includes: S(spiral), PM(Pre-Merger), M(Merger), LM(Later Stage of Merger) and E(Elliptical). After compare our classification with previous works, we find that our classification is consistent with those of others.

3) The fraction of spiral galaxies is lower in LIRGs compare to their counterparts in higher redshift. The lower $L_{IR}$ in local spiral galaxies also indicate that interaction between galaxies is need to induce a extremely $L_{IR}$ in local universe.

4) We also found that the advanced merging objects tend to have concentrated star-formation distribution , higher $L_{IR}$ and warmer far-IR color. All these results are consistent with the model that merger drive gas inward toward the nucleus and the star formation activity will be concentrated and enhanced as the merging process advances.

\acknowledgements


	   We are grateful to the referee for their careful reading of the paper and useful comments and suggestions. This project is supported by National Key R\&D Program of China (No.2017YFA0402704),
	and the National Natural Science Foundation of China (Grant Nos. 11733006,
	U1831119). The author acknowledge the hard work of observation assistant for 2.16 meter telescope: Hong-Bin Li, Jia-Ming Ai, Junjun Jia and Feng Xiao. This work was partially supported by the Open Project Program of the Key Laboratory of Optical Astronomy, National Astronomical Observatories, Chinese Academy of Sciences. This research has made use of the NASA/ IPAC Infrared Science Archive, which is operated by the Jet Propulsion Laboratory, California Institute of Technology, under contract with the National Aeronautics and Space Administratio. This work also used the published date from the Pan-STARRS1 Surveys (PS1).

\clearpage
\centering
	\tiny{
		\setlength{\tabcolsep}{3pt}
\begin{longtable*}[l]{lrrrrrrrrrrl}
	\caption{The main parameter and observation information of our source. }\\
	\hline 
	\hline 	
	(1)            & (2) & (3) & (4)    & (5)& (6) & (7) & (8) & (9)& (10) & (11)& (12)\\
	name              & RA. & Dec & cz    & Dist\footnote {For most objects this was calculated from cz using the cosmic attractor model \citep{2000ApJ...529..786M} with $H_{0}=75$ $km$ $s^{-1}$ $Mpc^{-1}$ and a flat universe where $\Omega_{M}=0.3$ and $\Omega_{\Lambda}=0.7$.} & F($25 \mu m$)    & F($60 \mu m$)  & F($100 \mu m$) & $L_{IR}$ \footnote {The log of the 8-1000 micron Luminosity determined using all IRAS bands with the units of solar bolometric luminosity ($L_{\sun}=3.83 \times 10^{33} srg$ $s^{-1}$). }  & Filter\footnote {$H\alpha$ filter used in observation.} &Date  & Seeing \\
	
	& J2000 & J2000 &  $km$ $s^{-1}$  & Mpc & Jy    & Jy & Jy & $L_{\sun}$ &   &    & arcsec\\

	\hline
	Arp220           & 15:34:57.12  & +23:30:11.5  & 5434  & 87.9  & 8.0   & 104.09 & 115.29 & 12.28 & c3  & 2007/05/15  & 2.7    \\
	CGCG 011-076     & 11:21:12.26  & -02:59:03.5  & 7464  & 117.0 & 0.76  & 5.85   & 9.18   & 11.43 & c4  & 2007/04/17  & 2.4    \\
	CGCG 043-099     & 13:01:50.80  & +04:20:00.0  & 11237 & 175.0 & 0.47  & 5.25   & 8.06   & 11.68 & c6  & 2008/02/13  & 2.1    \\
	CGCG 049-057     & 15:13:13.09  & +07:13:31.8  & 3897  & 65.4  & 0.95  & 21.89  & 31.53  & 11.35 & c3  & 2007/03/17  & 2.2    \\
	CGCG 052-037     & 16:30:56.54  & +04:04:58.4  & 7342  & 116.0 & 0.81  & 7.00    & 11.23  & 11.45 & c4  & 2008/04/03   & 2.0     \\
	CGCG 141-034     & 17:56:56.63  & +24:01:01.6  & 5944  & 93.4  & 0.56  & 6.24   & 10.55  & 11.20  & c4  & 2006/07/22  & 3.0      \\
	CGCG 142-034     & 18:16:40.66  & +22:06:46.1  & 5599  & 88.1  & 0.55  & 6.25   & 11.94  & 11.18 & c4  & 2006/07/21  & 2.2    \\
	CGCG 247-020     & 14:19:43.25  & +49:14:11.7  & 7716  & 120.0 & 0.84  & 6.01   & 8.47   & 11.39 & c4  & 2007/02/23  & 3.0      \\
	CGCG 436-030     & 01:20:02.72  & +14:21:42.9  & 9362  & 134.0 & 1.54  & 10.71  & 9.67   & 11.69 & c5  & 2007/09/10  & 2.5    \\
	CGCG 448-020     & 20:57:23.90  & +17:07:39.0  & 10822 & 161.0 & 2.30   & 12.65  & 11.76  & 11.94 & c6  & 2007/09/09   & 2.5    \\
	CGCG 453-062     & 23:04:56.53  & +19:33:08.0  & 7524  & 109.0 & 0.54  & 7.19   & 11.73  & 11.38 & c4  & 2006/09/25  & 3.2    \\
	CGCG 465-012     & 03:54:16.08  & +15:55:43.4  & 6662  & 94.3  & 0.75  & 5.65   & 8.95   & 11.20  & c4  & 2007/02/15  & 2.0      \\
	CGCG 468-002     & 05:08:20.50  & +17:21:58.0  & 5454  & 77.9  & 1.16  & 9.66   & 14.59  & 11.22 & c3  & 2008/02/10  & 2.1    \\
	ESO 453-G005     & 16:47:31.06  & -29:21:21.6  & 6260  & 100.0 & 0.61  & 9.56   & 12.17  & 11.37 & c4  & 2008/06/04   & 2.4    \\
	ESO 467-G027     & 22:14:39.92  & -27:27:50.3  & 5217  & 77.3  & 0.58  & 5.58   & 12.48  & 11.08 & c3  & 2007/11/10 & 2.1    \\
	ESO 507-G070     & 13:02:52.35  & -23:55:17.7  & 6506  & 106.0 & 0.80   & 13.04  & 15.71  & 11.56 & c4  & 2007/02/14  & 3.4    \\
	ESO 550-IG02(N)  & 04:21:20.02  & -18:48:47.6  & 9621  & 138.5 & 0.27  & 3.05   & 5.08   & 11.24 & c5  & 2007/02/14  & 3.4    \\
	ESO 550-IG02(S)  & 04:21:20.02  & -18:48:47.6  & 9621  & 138.5 & 0.24  & 2.64   & 4.39   & 11.18 & c5  & 2007/11/10 & 1.8    \\
	ESO 557-G002     & 06:31:47.22  & -17:37:17.3  & 6385  & 93.6  & 0.86  & 7.42   & 10.50   & 11.25 & c4  & 2007/02/14  & 2.2    \\
	ESO 593-IG008    & 19:14:30.90  & -21:19:07.0  & 14608 & 222.0 & 0.51  & 6.38   & 9.37   & 11.93 & c7  & 2008/06/07   & 2.4    \\
	ESO 602-G025     & 22:31:25.48  & -19:02:04.1  & 7507  & 110.0 & 0.91  & 5.42   & 9.64   & 11.34 & c4  & 2007/10/11 & 2.6    \\
	IC 0214          & 02:14:05.59  & +05:10:23.7  & 9061  & 129.0 & 0.62  & 5.28   & 8.57   & 11.43 & c5  & 2007/11/10 & 2.4    \\
	IC 0860          & 13:15:03.53  & +24:37:07.9  & 3347  & 56.8  & 1.34  & 18.61  & 18.66  & 11.14 & c3  & 2007/02/14  & 2.1    \\
	IC 1623A         & 01:07:47.18  & -17:30:25.3  & 6016  & 85.5  & 3.65  & 22.93  & 31.55  & 11.71 & c4  & 2007/11/11 & 2.5    \\
	IC 2810          & 11:25:47.30  & +14:40:21.1  & 10192 & 157.0 & 0.62  & 6.20    & 10.39  & 11.64 & c6  & 2007/02/15  & 2.2    \\
	IC 4280          & 13:32:53.40  & -24:12:25.7  & 4889  & 82.4  & 0.68  & 6.10    & 12.36  & 11.15 & c3  & 2008/02/07   & 4.2    \\
	IC 5298          & 23:16:00.70  & +25:33:24.1  & 8221  & 119.0 & 1.95  & 9.06   & 11.99  & 11.60  & c5  & 2008/09/29  & 1.8    \\
	IC 563           & 09:46:20.71  & +03:03:30.5  & 5996  & 92.9  & 0.27  & 2.68   & 6.18   & 10.94 & c4  & 2007/04/17  & 2.1    \\
	IC 564           & 09:46:20.71  & +03:03:30.5  & 5996  & 92.9  & 0.27  & 2.59   & 6.00    & 10.92 & c4  & 2007/04/17  & 2.1    \\
	III Zw 035(N)    & 01:44:30.45  & +17:06:05.0  & 8375  & 119.0 & 0.52  & 6.63   & 7.15   & 11.3  & c5  & 2007/09/10  & 2.1    \\
	III Zw 035(S)    & 01:44:30.45  & +17:06:05.0  & 8375  & 119.0 & 0.51  & 6.62   & 7.15   & 11.3  & c5  & 2007/09/10  & 2.1    \\
	IRAS 03582+6012  & 04:02:32.48  & +60:20:40.1  & 8997  & 131.0 & 0.70   & 5.65   & 7.76   & 11.43 & c5  & 2007/11/12 & 2.1    \\
	IRAS 04271+3849  & 04:30:33.09  & +38:55:47.7  & 5640  & 80.8  & 0.74  & 6.53   & 10.36  & 11.11 & c4  & 2008/02/07   & 2      \\
	IRAS 05083+2441  & 05:11:25.88  & +24:45:18.3  & 6915  & 99.2  & 0.85  & 6.92   & 8.36   & 11.26 & c4  & 2007/02/23  & 2.4    \\
	IRAS 05129+5128  & 05:16:56.10  & +51:31:56.5  & 8224  & 120.0 & 1.05  & 6.56   & 7.34   & 11.42 & c5  & 2007/02/23  & 2.3    \\
	IRAS 05442+1732  & 05:47:11.18  & +17:33:46.7  & 5582  & 80.5  & 1.70   & 10.02  & 12.73  & 11.30  & c4  & 2007/02/14  & 1.6    \\
	IRAS 17578-0400  & 18:00:31.90  & -04:00:53.3  & 4210  & 68.5  & 1.14  & 27.69  & 33.1   & 11.48 & c3  & 2007/10/10 & 3.3    \\
	IRAS 18090+0130  & 18:11:35.91  & +01:31:41.3  & 8662  & 134.0 & 0.81  & 7.73   & 15.64  & 11.65 & c5  & 2007/10/11 & 2.0      \\
	IRAS 19542+1110  & 19:56:35.44  & +11:19:02.6  & 19473 & 295.0 & 0.77  & 6.18   & 6.22   & 12.12 & c10 & 2008/06/06   & 3.0      \\
	IRAS 20351+2521  & 20:37:17.72  & +25:31:37.7  & 10102 & 151.0 & 0.71  & 5.93   & 8.95   & 11.61 & c5  & 2007/09/10  & 3.0      \\
	IRAS 21101+5810  & 21:11:30.40  & +58:23:03.2  & 11705 & 174.0 & 0.82  & 7.94   & 11.08  & 11.81 & c6  & 2008/09/29  & 2.0      \\
	IRAS 23436+5257  & 23:46:05.58  & +53:14:00.6  & 10233 & 149.0 & 0.74  & 5.66   & 9.01   & 11.57 & c6  & 2008/09/29  & 1.7    \\
	IRAS F01364-1042 & 01:38:52.92  & -10:27:11.4  & 14464 & 210.0 & 0.44  & 6.62   & 6.88   & 11.85 & c7  & 2008/09/28  & 3.3    \\
	IRAS F02437+2122 & 02:46:39.15  & +21:35:10.3  & 6987  & 98.8  & 0.63  & 5.90    & 6.67   & 11.16 & c4  & 2007/09/09   & 2.5    \\
	IRAS F03217+4022 & 03:25:05.38  & +40:33:29.0  & 7007  & 100.0 & 0.91  & 7.47   & 10.87  & 11.33 & c4  & 2007/11/12 & 2.1    \\
	IRAS F03359+1523 & 03:38:46.70  & +15:32:55.0  & 10613 & 152.0 & 0.65  & 5.97   & 7.27   & 11.55 & c6  & 2007/11/10 & 1.5    \\
	IRAS F05187-1017 & 05:21:06.54  & -10:14:46.7  & 8474  & 122.0 & 0.19  & 5.39   & 8.04   & 11.30  & c5  & 2007/03/17  & 2.5    \\
	IRAS F05189-2524 & 05:21:01.47  & -25:21:45.4  & 12760 & 187.0 & 3.47  & 13.25  & 11.84  & 12.16 & c7  & 2008/02/14  & 2.5    \\
	IRAS F06076-2139 & 06:09:45.81  & -21:40:23.7  & 11226 & 165.0 & 0.63  & 6.43   & 8.47   & 11.65 & c6  & 2008/02/08   & 3.0   \\
	IRAS F08339+6517 & 08:38:23.18  & +65:07:15.2  & 5730  & 86.3  & 1.13  & 5.81   & 6.48   & 11.11 & c4  & 2007/11/06  & 1.5    \\
	IRAS F08572+3915 & 09:00:25.39  & +39:03:54.4  & 17493 & 264.0 & 1.76  & 7.30    & 4.77   & 12.16 & c9  & 2008/02/13  & 1.5    \\
	IRAS F09111-1007 & 09:13:37.61  & -10:19:24.8  & 16231 & 246.0 & 0.74  & 6.75   & 10.68  & 12.06 & c8  & 2008/02/13  & 1.5    \\
	IRAS F10173+0828 & 10:20:00.21  & +08:13:33.8  & 14716 & 224.0 & 0.55  & 5.61   & 5.86   & 11.86 & c7  & 2008/02/14  & 1.8    \\
	IRAS F10565+2448 & 10:59:18.14  & +24:32:34.3  & 12921 & 197.0 & 1.27  & 12.10   & 15.01  & 12.08 & c7  & 2008/02/14  & 1.8    \\
	IRAS F12112+0305 & 12:13:46.00  & +02:48:38.0  & 21980 & 340.0 & 0.66  & 8.18   & 9.46   & 12.36 & c11 & 2008/02/13  & 1.8    \\
	IRAS F12224-0624 & 12:25:03.91  & -06:40:52.6  & 7902  & 125.0 & 0.20   & 5.99   & 8.13   & 11.36 & c5  & 2008/01/31  & 2.3    \\
	IRAS F15250+3608 & 15:26:59.40  & +35:58:37.5  & 16535 & 254.0 & 1.31  & 7.10    & 5.93   & 12.08 & c8  & 2008/02/13  & 2.7    \\
	IRAS F16164-0746 & 16:19:11.79  & -07:54:02.8  & 8140  & 128.0 & 0.59  & 10.29  & 13.22  & 11.62 & c5  & 2008/02/09   & 3.3    \\
	IRAS F16399-0937 & 16:42:40.21  & -09:43:14.4  & 8098  & 128.0 & 1.13  & 8.42   & 14.72  & 11.63 & c5  & 2008/02/09   & 2.7    \\
	IRAS F16516-0948 & 16:54:24.03  & -09:53:20.9  & 6755  & 107.0 & 0.49  & 5.32   & 11.65  & 11.31 & c4  & 2008/05/02   & 2.1    \\
	IRAS F17132+5313 & 17:14:20.00  & +53:10:30.0  & 15270 & 232.0 & 0.65  & 6.07   & 7.90    & 11.96 & c8  & 2007/11/12 & 1.8    \\
	IRAS F17138-1017 & 17:16:35.79  & -10:20:39.4  & 5197  & 84.0  & 2.12  & 15.18  & 19.02  & 11.49 & c3  & 2007/09/08   & 1.8    \\
	IRAS F17207-0014 & 17:23:21.95  & -00:17:00.9  & 12834 & 198.0 & 1.61  & 32.13  & 36.08  & 12.46 & c7  & 2008/05/02   & 2.6    \\
	IRAS F22491-1808 & 22:51:49.26s  & -17:52:23.5  & 23312 & 351.0 & 0.54  & 5.54  & 4.64   & 12.20  & c11 & 2008/09/28  & 4.0      \\
	IRAS F23365+3604 & 23:39:01.27  & +36:21:08.7  & 19331 & 287.0 & 0.94  & 7.44   & 9.01   & 12.20  & c10 & 2007/11/11 & 2.5    \\
	MCG+02-20-003    & 07:35:43.37  & +11:42:33.5  & 4873  & 72.8  & 0.81  & 9.38   & 13.33  & 11.13 & c3  & 2009/01/17  & 2.5    \\
	MCG+04-48-002    & 20:28:35.06  & +25:44:00.0  & 4167  & 64.2  & 1.09  & 9.93   & 17.36  & 11.11 & c3  & 2006/09/23  & 1.6    \\
	MCG+05-06-036    & 02:23:21.99  & +32:11:49.5  & 10106 & 145.0 & 0.80   & 6.55   & 11.63  & 11.64 & c5  & 2007/11/11 & 2.5    \\
	MCG+07-23-019    & 11:03:53.20  & +40:50:57.0  & 10350 & 158.0 & 0.71  & 6.38   & 10.30   & 11.62 & c6  & 2007/02/23  & 2.5    \\
	MCG+08-11-002    & 05:40:43.71  & +49:41:41.5  & 5743  & 83.7  & 1.08  & 14.03  & 24.82  & 11.46 & c4  & 2007/02/14  & 1.7    \\
	MCG+08-18-013    & 09:36:37.19  & +48:28:27.7  & 7777  & 117.0 & 0.75  & 5.68   & 8.42   & 11.34 & c4  & 2008/02/06   & 2.4    \\
	MCG+12-02-001    & 00:54:03.61  & +73:05:11.8  & 4706  & 69.8  & 3.51  & 21.92  & 29.11  & 11.50  & c3  & 2007/09/08   & 1.8    \\
	MCG-02-01-051    & 00:18:50.51  & -10:22:09.2  & 8159  & 117.5 & 0.66  & 4.11   & 5.31   & 11.22 & c5  & 2007/10/10 & 3.0      \\
	MCG-02-01-052    & 00:18:50.51  & -10:22:09.2  & 8159  & 117.5 & 0.54  & 3.37   & 4.35   & 11.13 & c5  & 2007/10/10 & 3.0      \\
	MCG-02-33-098    & 13:02:19.70  & -15:46:03.0  & 4773  & 78.7  & 1.63  & 7.49   & 9.68   & 11.17 & c3  & 2007/02/15  & 2.0      \\
	MCG-03-04-014    & 01:10:08.96  & -16:51:09.8  & 10040 & 144.0 & 0.90   & 7.25   & 10.33  & 11.65 & c6  & 2007/11/10 & 1.8    \\
	MCG-03-34-064    & 13:22:24.46  & -16:43:42.9  & 4959  & 82.2  & 2.97  & 6.20    & 6.20    & 11.28 & c3  & 2008/02/07   & 3.0      \\
	MCG01-60-022     & 23:42:00.85  & -03:36:54.6  & 6966  & 100.0 & 1.06  & 5.39   & 8.26   & 11.27 & c4  & 2007/09/09   & 3.0      \\
	Mrk231           & 12:56:14.24  & +56:52:25.2  & 12642 & 192.0 & 8.84  & 30.80   & 29.74  & 12.57 & c6  & 2007/04/17  & 2.4    \\
	Mrk331           & 23:51:26.80  & +20:35:09.9  & 5541  & 79.3  & 2.54  & 18.00   & 22.70   & 11.50  & c3  & 2007/09/08   & 1.8    \\
	NGC 0317B        & 00:57:40.45  & +43:47:32.1  & 5429  & 77.8  & 1.03  & 9.16   & 13.60   & 11.19 & c3  & 2007/09/08   & 1.8    \\
	NGC 0838         & 02:09:38.58  & -10:08:46.3  & 3851  & 53.8  & 1.88  & 11.41  & 19.94  & 11.05 & c3  & 2007/09/08   & 1.8    \\
	NGC 0958         & 02:30:42.83  & -02:56:20.4  & 5738  & 80.6  & 0.94  & 5.85   & 15.08  & 11.20  & c4  & 2007/02/15  & 2.5    \\
	NGC 0992         & 02:37:25.49  & +21:06:03.0  & 4141  & 58.0  & 1.76  & 11.40   & 16.72  & 11.07 & c3  & 2007/02/14  & 1.7    \\
	NGC 1068         & 02:42:40.71  & -00:00:47.8  & 1137  & 15.9  & 87.57 & 196.37 & 257.37 & 11.40  & c1  & 2007/02/14  & 2.1    \\
	NGC 1275         & 03:19:48.16  & +41:30:42.1  & 5264  & 75.0  & 3.44  & 6.99   & 7.2    & 11.26 & c3  & 2009/01/29  & 2.0      \\
	NGC 1614         & 04:33:59.85  & -08:34:44.0  & 4778  & 67.8  & 7.50   & 32.12  & 34.32  & 11.65 & c3  & 2007/11/10 & 1.5    \\
	NGC 1797         & 05:07:44.88  & -08:01:08.7  & 4441  & 63.4  & 1.35  & 9.56   & 12.76  & 11.04 & c3  & 2008/02/07   & 2.1    \\
	NGC 1961         & 05:42:04.65  & +69:22:42.4  & 3934  & 59.0  & 0.99  & 7.17   & 23.37  & 11.06 & c3  & 2007/10/11 & 2.0      \\
	NGC 2146         & 06:18:37.71  & +78:21:25.3  & 893   & 17.5  & 18.81 & 146.69 & 194.05 & 11.12 & c1  & 2007/03/16  & 2.5    \\
	NGC 23           & 00:09:53.41  & +25:55:25.6  & 4566  & 65.2  & 1.29  & 9.03   & 15.66  & 11.12 & c3  & 2006/09/22  & 2.0      \\
	NGC 232          & 00:42:45.82  & -23:33:40.9  & 6647  & 95.2  & 1.28  & 10.05  & 17.14  & 11.44 & c4  & 2007/10/10 & 2.5    \\
	NGC 2342         & 07:09:18.08  & +20:38:09.5  & 5276  & 78.0  & 1.64  & 7.73   & 24.10   & 11.31 & c3  & 2007/03/13  & 1.9    \\
	NGC 2388         & 07:28:53.44  & +33:49:08.7  & 4134  & 62.1  & 1.98  & 16.74  & 24.58  & 11.28 & c3  & 2006/02/23  & 3.9    \\
	NGC 2623         & 08:38:24.08  & +25:45:16.6  & 5549  & 84.1  & 1.81  & 23.74  & 25.88  & 11.60  & c3  & 2007/03/17  & 2.3    \\
	NGC 3110         & 10:04:02.11  & -06:28:29.2  & 5054  & 79.5  & 1.13  & 11.28  & 22.27  & 11.37 & c3  & 2007/03/13  & 2.3    \\
	NGC 3221         & 10:22:19.98  & +21:34:10.5  & 4110  & 65.7  & 0.93  & 7.72   & 18.76  & 11.09 & c3  & 2006/02/22  & 2.6    \\
	NGC 34           & 00:11:06.55  & -12:06:26.3  & 5881  & 84.1  & 2.39  & 17.05  & 16.86  & 11.49 & c4  & 2007/10/11 & 2.8    \\
	NGC 3690(E)    & 11:28:32.25  & +58:33:44.0  & 3093  & 50.7  & 11.98 & 55.28  & 54.48  & 11.62 & c2  & 2007/03/17  & 2.0      \\
	NGC 3690(W)    & 11:28:32.25  & +58:33:44.0  & 3093  & 50.7  & 12.53 & 57.77  & 56.94  & 11.64 & c2  & 2007/03/17  & 2.0      \\
	NGC 4194         & 12:14:09.47  & +54:31:36.6  & 2501  & 43.0  & 4.51  & 23.20   & 25.16  & 11.10  & c2  & 2008/02/09   & 2.1    \\
	NGC 4418         & 12:26:54.62  & -00:52:39.2  & 2179  & 36.5  & 9.67  & 43.89  & 31.94  & 11.19 & c2  & 2007/02/14  & 2.3    \\
	NGC 4922         & 13:01:24.89  & +29:18:40.0  & 7071  & 111.0 & 1.48  & 6.21   & 7.33   & 11.38 & c4  & 2008/02/09   & 2.4    \\
	NGC 5104         & 13:21:23.08  & +00:20:32.7  & 5578  & 90.8  & 0.74  & 6.78   & 13.37  & 11.27 & c4  & 2007/02/23  & 2.8    \\
	NGC 5135         & 13:25:44.06  & -29:50:01.2  & 4105  & 60.9  & 2.38  & 16.86  & 30.97  & 11.30  & c3  & 2008/02/12  & 4.2    \\
	NGC 5256         & 13:38:17.52  & +48:16:36.7  & 8341  & 129.0 & 1.07  & 7.25   & 10.11  & 11.56 & c5  & 2007/02/24  & 1.9    \\
	NGC 5257         & 13:39:55.00  & +00:50:07.0  & 6778  & 108.5 & 0.76  & 6.11   & 11.36  & 11.38 & c4  & 2007/03/17  & 1.9    \\
	NGC 5258         & 13:39:55.00  & +00:50:07.0  & 6778  & 108.5 & 0.58  & 4.62   & 8.61   & 11.25 & c4  & 2007/03/17  & 1.9    \\
	NGC 5331         & 13:52:16.29  & +02:06:17.0  & 9906  & 155.0 & 0.59  & 5.86   & 11.49  & 11.66 & c5  & 2007/03/14  & 2.2    \\
	NGC 5394         & 13:58:35.81  & +37:26:20.3  & 3482  & 58.7  & 0.61  & 4.09   & 9.59   & 10.72 & c3  & 2007/02/14  & 2.1    \\
	NGC 5395         & 13:58:35.81  & +37:26:20.3  & 3482  & 58.7  & 0.79  & 5.30    & 12.43  & 10.83 & c3  & 2007/02/14  & 2.1    \\
	NGC 5653         & 14:30:10.42  & +31:12:55.8  & 3562  & 60.2  & 1.37  & 10.57  & 23.03  & 11.13 & c3  & 2007/02/14  & 2.3    \\
	NGC 5734         & 14:45:09.05  & -20:52:13.7  & 4121  & 67.1  & 0.74  & 7.99   & 24.79  & 11.15 & c3  & 2007/03/17  & 2.9    \\
	NGC 5936         & 15:30:00.84  & +12:59:21.5  & 4004  & 67.1  & 1.47  & 8.73   & 17.66  & 11.14 & c3  & 2006/07/2   & 2.1    \\
	NGC 5990         & 15:46:16.37  & +02:24:55.7  & 3839  & 64.4  & 1.60   & 9.59   & 17.14  & 11.13 & c3  & 2007/03/16  & 1.8    \\
	NGC 6052         & 16:05:13.05  & +20:32:32.6  & 4739  & 77.6  & 0.83  & 6.79   & 10.57  & 11.09 & c3  & 2006/07/5   & 3.9    \\
	NGC 6090         & 16:11:40.70  & +52:27:24.0  & 8947  & 137.0 & 1.24  & 6.48   & 9.41   & 11.58 & c5  & 2008/05/02   & 1.8    \\
	NGC 6240         & 16:52:58.89  & +02:24:03.4  & 7339  & 116.0 & 3.55  & 22.94  & 26.49  & 11.93 & c4  & 2008/02/08   & 5.0      \\
	NGC 6286         & 16:58:31.38  & +58:56:10.5  & 5501  & 85.7  & 0.62  & 9.24   & 23.11  & 11.37 & c4  & 2006/07/02   & 2.9    \\
	NGC 6621         & 18:12:55.31  & +68:21:48.4  & 6191  & 94.3  & 0.97  & 6.78   & 12.01  & 11.29 & c4  & 2007/05/15  & 2.5    \\
	NGC 6670(E)      & 18:33:35.91  & +59:53:20.2  & 8574  & 129.5 & 0.52  & 4.46   & 7.05   & 11.35 & c5  & 2006/07/22  & 2.4    \\
	NGC 6670(W)      & 18:33:35.91  & +59:53:20.2  & 8574  & 129.5 & 0.53  & 4.52   & 7.14   & 11.35 & c5  & 2006/07/22  & 2.4    \\
	NGC 6701         & 18:43:12.46  & +60:39:12.0  & 3965  & 62.4  & 1.32  & 10.05  & 20.05  & 11.12 & c3  & 2006/07/22  & 2.7    \\
	NGC 6786         & 19:10:59.20  & +73:25:06.3  & 7528  & 113.0 & 1.42  & 7.58   & 10.77  & 11.49 & c4  & 2006/09/22  & 1.9    \\
	NGC 6907         & 20:25:06.65  & -24:48:33.5  & 3190  & 50.1  & 1.94  & 14.14  & 29.59  & 11.11 & c2  & 2007/11/11 & 2.5    \\
	NGC 6926         & 20:33:06.11  & -02:01:39.0  & 5880  & 89.1  & 1.03  & 7.09   & 14.38  & 11.32 & c4  & 2006/09/23  & 1.9    \\
	NGC 695          & 01:51:14.24  & +22:34:56.5  & 9735  & 139.0 & 0.83  & 7.59   & 13.56  & 11.68 & c5  & 2007/09/10  & 2.1    \\
	NGC 7469         & 23:03:15.62  & +08:52:26.4  & 4892  & 70.8  & 5.96  & 27.33  & 35.16  & 11.65 & c3  & 2007/09/08   & 1.8    \\
	NGC 7591         & 23:18:16.28  & +06:35:08.9  & 4956  & 71.4  & 1.27  & 7.87   & 14.87  & 11.12 & c3  & 2006/09/22  & 2.3    \\
	NGC 7592(E)      & 23:18:22.20  & -04:24:57.6  & 7328  & 106.0 & 0.55  & 4.60    & 6.04   & 11.16 & c4  & 2007/09/09   & 3.0      \\
	NGC 7592(W)      & 23:18:22.20  & -04:24:57.6  & 7328  & 106.0 & 0.42  & 3.46   & 4.54   & 11.03 & c4  & 2007/09/09   & 3.0     \\
	NGC 7674         & 23:27:56.72  & +08:46:44.5  & 8671  & 125.0 & 1.92  & 5.36   & 8.33   & 11.56 & c5  & 2007/09/10  & 3.0      \\
	NGC 7679         & 23:28:46.66  & +03:30:41.1  & 5138  & 73.8  & 1.12  & 7.40    & 10.71  & 11.11 & c3  & 2007/09/08   & 1.5    \\
	NGC 7752         & 23:47:01.70  & +29:28:16.3  & 5120  & 73.6  & 0.46  & 3.20    & 6.40    & 10.81 & c3  & 2008/09/29  & 1.7    \\
	NGC 7753         & 23:47:01.70  & +29:28:16.3  & 5120  & 73.6  & 0.37  & 2.59   & 5.18   & 10.72 & c3  & 2008/09/29  & 1.7    \\
	NGC 7771         & 23:51:24.88  & +20:06:42.6  & 4277  & 61.2  & 2.17  & 19.67  & 40.12  & 11.40  & c3  & 2006/09/22  & 2.1    \\
	NGC 828          & 02:10:09.57  & +39:11:25.3  & 5374  & 76.3  & 1.07  & 11.46  & 25.33  & 11.36 & c3  & 2006/09/22  & 2.0      \\
	NGC 877          & 02:17:59.64  & +14:32:38.6  & 3913  & 54.6  & 1.41  & 11.82  & 25.56  & 11.10  & c3  & 2006/09/23  & 2.0      \\
	UGC 01385        & 01:54:53.79  & +36:55:04.6  & 5621  & 79.8  & 0.99  & 5.89   & 7.81   & 11.05 & c4  & 2007/02/15  & 2      \\
	UGC 03410        & 06:14:29.63  & +80:26:59.6  & 3921  & 59.7  & 0.92  & 9.87   & 22.98  & 11.10 & c3  & 2009/01/18  & 1.8    \\
	UGC 04881        & 09:15:55.11  & +44:19:54.1  & 11851 & 178.0 & 0.61  & 6.07   & 10.33  & 11.74 & c6  & 2007/02/15  & 2.3    \\
	UGC 08387        & 13:20:35.34  & +34:08:22.2  & 6985  & 110.0 & 1.42  & 17.04  & 24.38  & 11.73 & c4  & 2007/02/15  & 2      \\
	UGC 08739        & 13:49:13.93  & +35:15:26.8  & 5032  & 81.4  & 0.42  & 5.79   & 15.89  & 11.15 & c3  & 2007/02/15  & 1.4    \\
	UGC 11041        & 17:54:51.82  & +34:46:34.4  & 4881  & 77.5  & 0.69  & 5.84   & 12.78  & 11.11 & c3  & 2006/07/05   & 3.5    \\
	UGC 12150        & 22:41:12.26  & +34:14:57.0  & 6413  & 93.5  & 0.82  & 8.00    & 15.58  & 11.35 & c4  & 2006/07/22  & 2.2    \\
	UGC 1845         & 02:24:07.98  & +47:58:11.0  & 4679  & 67.0  & 1.07  & 10.31  & 15.51  & 11.12 & c3  & 2007/10/11 & 2.4    \\
	UGC 2238         & 02:46:17.49  & +13:05:44.4  & 6560  & 92.4  & 0.65  & 8.17   & 15.67  & 11.33 & c4  & 2006/09/22  & 2.3    \\
	UGC 2369         & 02:54:01.78  & +14:58:24.9  & 9558  & 136.0 & 1.88  & 8.07   & 11.18  & 11.67 & c5  & 2006/09/23  & 2.0      \\
	UGC 2608         & 03:15:01.42  & +42:02:09.4  & 6998  & 100.0 & 1.45  & 8.18   & 11.27  & 11.41 & c4  & 2008/02/06   & 2.1    \\
	UGC 2982         & 04:12:22.45  & +05:32:50.6  & 5305  & 74.9  & 0.83  & 8.39   & 16.82  & 11.20  & c3  & 2007/10/11 & 3.0      \\
	UGC 3094         & 04:35:33.83  & +19:10:18.2  & 7408  & 106.0 & 0.84  & 6.35   & 12.85  & 11.41 & c4  & 2007/11/06  & 1.4    \\
	UGC 3351         & 05:45:47.88  & +58:42:03.9  & 4455  & 65.8  & 0.86  & 14.26  & 29.46  & 11.28 & c3  & 2008/09/29  & 1.8    \\
	UGC 3608         & 06:57:34.45  & +46:24:10.8  & 6401  & 94.3  & 1.20   & 8.05   & 11.33  & 11.34 & c4  & 2007/11/06  & 1.5    \\
	UGC 5101         & 09:35:51.65  & +61:21:11.3  & 11802 & 177.0 & 1.02  & 11.68  & 19.91  & 12.01 & c6  & 2008/02/07   & 2.4    \\
	VII ZW 031       & 05:16:46.44  & +79:40:12.6  & 16090 & 240.0 & 0.62  & 5.51   & 10.09  & 11.99 & c8  & 2007/11/12 & 2.5    \\
	VV250a           & 13:15:35.06  & +62:07:28.6  & 9313  & 142.0 & 1.95  & 11.39  & 12.41  & 11.81 & c5  & 2007/05/16  & 1.8    \\
	VV340a           & 14:57:00.68  & +24:37:02.7  & 10094 & 157.0 & 0.41  & 6.95   & 15.16  & 11.74 & c5  & 2007/07/01   & 2.1    \\
	VV705            & 15:18:06.28  & +42:44:41.2  & 11944 & 183.0 & 1.42  & 9.02   & 10.00   & 11.92 & c6  & 2007/04/17  & 2.1     \\

	\hline
	\hline
	\label{table:RBGS}
	
\end{longtable*}
}

\clearpage
\begin{longtable*}[l]{llrrrrr}
	\caption{The H$\alpha$ luminosity information. }\\
	\hline 
	\hline 
	\hline	
	(1)               & (2) & (3)& (4)&(5)&(6)&(7)\\
	name                  & type\footnote {The morphology and interactions type of (U-)LIRGs and the unknown objects is masked by $*$.} & $logL(H\alpha)_{obs}$\footnote{The $H\alpha$ luminosity after corrected for transmission, [NII] emission and Galactic extinction.} & $logL(H\alpha)$\footnote{The $H\alpha$ luminosity after corrected for internal extinction.} & $Frac(H\alpha)$\footnote{The ratio between flux enclosed $R_{e}$ and total.} &$PA$&$e$\\
	& & $erg$ $s^{-1}$ $cm^{-2}$& $erg$ $s^{-1}$ $cm^{-2}$& &degree\\
	\hline
	Arp220           & M                         & 41.76   & 43.20   & 0.69     & 20.00   & 0.10     \\
	CGCG 011-076     & S                         & 41.87   & 42.52   & 0.77     & -82.57  & 0.38     \\
	CGCG 043-099     & M                         & 41.33   & 42.56   & 0.74     & -64.31  & 0.15     \\
	CGCG 049-057     & $E^{*}$  & 40.71   & 42.02   & 1.00    & -14.80  & 0.07     \\
	CGCG 052-037     & LM                        & 42.09   & 42.59   & 0.75     & 52.07   & 0.27     \\
	CGCG 141-034     & $S^{*}$ & 41.35   & 42.15   & 0.88     & 23.90   & 0.13     \\
	CGCG 142-034     & M                         & 41.47   & 42.13   & 0.63     & 74.97   & 0.64     \\
	CGCG 247-020     & E                         & 41.63   & 42.52   & 0.78     & -55.31  & 0.02     \\
	CGCG 436-030     & PM                         & 42.00   & 42.90   & 0.82     & 32.29   & 0.19     \\
	CGCG 448-020     & M                         & 42.80   & 43.31   &0.60     & 47.64   & 0.53     \\
	CGCG 453-062     & $LM^{*}$ & 42.08   & 42.44   & 0.45     & 85.34   & 0.08     \\
	CGCG 465-012     & M                         & 42.17   & 42.50   & 0.85     & 66.00   & 0.04     \\
	CGCG 468-002     & $M^{*}$  & 41.90   & 42.37   &0.65     & 25.00   & 0.60     \\
	ESO 453-G005     &$ M^{*}$  & 41.83   & 42.35   & 0.71     & -47.20  & 0.02     \\
	ESO 507-G070     & $M^{*}$  & 41.61   & 42.42   & 0.86     & -30.90  & 0.21     \\
	ESO 550-IG02(N)  & PM                         & 41.73   & 42.47   & 0.59     & 74.74   & 0.20     \\
	ESO 550-IG02(S)  & PM                         & 41.14   & 42.41   & 0.64     & -64.05  & 0.45     \\
	ESO 557-G002     & PM                         & 42.05   & 42.48   & 0.83     & -149.35 & 0.02     \\
	ESO 593-IG008    & M                         & 42.68   & 43.03   & 0.61     & 55.25   & 0.20     \\
	ESO 602-G025     & M                         & 41.92   & 42.55   & 0.81     & 9.36    & 0.69     \\
	ESO\_467-G027    & S                         & 42.02   & 42.28   & 0.74     & -52.76  & 0.34     \\
	IC 0214          & M                         & 42.32   & 42.67   &0.52     & 5.00    & 0.10     \\
	IC 0860          & $S^{*}$ & 40.31   & 42.15   & 0.63     & -18.09  & 0.12     \\
	IC 1623A         & M                         & 42.84   & 43.13   & 0.67     & -88.65  & 0.08     \\
	IC 2810          & S                         & 41.92   & 42.66   & 0.83     & -40.78  & 0.52     \\
	IC 4280          & S                         & 42.16   & 42.41   & 0.80     & -176.83 & 0.12     \\
	IC 5298          & M                         & 41.92   & 42.88   & 0.66     & -21.05  & 0.03     \\
	IC 563           & PM                         & 41.99   & 42.34   & 0.67     & 53.57   & 0.67     \\
	IC 564           & PM                         & 41.96   & 42.33   & 0.54     & -69.80  & 0.65     \\
	III Zw 035(N)    & $M^{*}$  & 41.20   & 42.57   &0.56     & -18.90  & 0.45     \\
	III Zw 035(S)    & $M^{*}$  & 41.20   & 42.57   & 0.36     & -44.50  & 0.21     \\
	IRAS 03582+6012  & $M^{*}$  & 42.27   & 42.69   & 0.66     & -59.95  & 0.65     \\
	IRAS 04271+3849  & LM                        & 41.92   & 42.31   &0.69     & 48.92   & 0.44     \\
	IRAS 05083+2441  & LM                        & 42.12   & 42.54   & 0.66     & -22.06  & 0.44     \\
	IRAS 05129+5128  & M                         & 42.23   & 42.74   & 0.81     & 81.78   & 0.42     \\
	IRAS 05442+1732  & LM                        & 41.98   & 42.58   & 0.84     & 73.82   & 0.31     \\
	IRAS 17578-0400  & $LM^{*}$ & 41.63   & 42.20   & 0.73     & 37.88   & 0.60     \\
	IRAS 18090+0130  & M                         & 42.26   & 42.73   & 0.56     & -14.18  & 0.47     \\
	IRAS 19542+1110  & E                         & 41.83   & 43.20   & 0.76     & -72.24  & 0.07     \\
	IRAS 20351+2521  & M                         & 42.04   & 42.70   & 0.96     & 10.15   & 0.09     \\
	IRAS 21101+5810  & M                         & 41.71   & 42.81   & 0.96     & -110.33 & 0.22     \\
	IRAS 23436+5257  & M                         & 42.10   & 42.72   & 0.81     & 22.89   & 0.62     \\
	IRAS F01364-1042 & E                         & 41.31   & 42.69   & 0.88     & -64.14  & 0.25     \\
	IRAS F02437+2122 & $E^{*}$  & 40.99   & 42.22   & 0.94     & 71.27   &0.36     \\
	IRAS F03217+4022 & LM                        & 41.17   & 42.15   &0.56     & 72.34   & 0.28     \\
	IRAS F03359+1523 & M                         & 42.24   & 42.73   & 0.72     & -74.38  & 0.73     \\
	IRAS F05187-1017 & $S^{*}$ & 41.17   & 41.93   & 0.56     & 70.99   & 0.35     \\
	IRAS F05189-2524 & E                         & 42.24   & 43.49   & 0.64     & -107.75 & 0.07     \\
	IRAS F06076-2139 & M                         & 41.80   & 42.68   & 0.63     & -13.06  & 0.22     \\
	IRAS F08339+6517 & LM                        & 42.58   & 42.77   & 0.67     & -153.41 & 0.05     \\
	IRAS F08572+3915 & M                         & 41.45   & 43.46   &0.91     & 57.36   & 0.43     \\
	IRAS F09111-1007 & PM                         & 42.33   & 43.10   & 0.68     & 22.79   & 0.07     \\
	IRAS F10173+0828 & $S^{*}$ & 41.56   & 42.84   & 0.75     & 76.02   & 0.54     \\
	IRAS F10565+2448 & M                         & 42.13   & 43.12   & 0.84     & 85.07   & 0.24     \\
	IRAS F12112+0305 & M                         & 42.09   & 43.26   & 0.76     & 54.77   & 0.22     \\
	IRAS F12224-0624 & $S^{*}$ & 40.60   & 41.90   & 0.60     & -32.78  & 0.16     \\
	IRAS F15250+3608 & M                         & 41.72   & 43.30   & 0.75     & -56.70  & 0.03     \\
	IRAS F16164-0746 & LM                        & 41.24   & 42.40   & 1.00   & 74.64   & 0.47     \\
	IRAS F16399-0937 & $M^{*}$  & 42.00   & 42.74   &0.62     & 4.27    & 0.60     \\
	IRAS F16516-0948 & M                         & 42.37   & 42.57   &0.50     & 20.00   & 0.10     \\
	IRAS F17132+5313 & $PM^{*}$  & 42.39   & 43.02   & 0.87     & 0.56     & 0.41     \\
	IRAS F17138-1017 & M                         & 41.96   & 42.66   & 0.85     & -1.53   & 0.43     \\
	IRAS F17207-0014 & LM                        & 42.11   & 43.21   & 0.59     & 56.78   & 0.16     \\
	IRAS F22491-1808 & LM                        & 42.23   & 43.22   & 0.65     & 78.85   & 0.24     \\
	IRAS F23365+3604 & LM                        & 42.03   & 43.28   &0.77     & -75.00  & 0.30     \\
	MCG+02-20-003    & M                         & 41.64   & 42.19   & 0.93     & 48.86   & 0.49     \\
	MCG+04-48-002    & M                         & 41.74   & 42.23   & 0.69     & -60.56  & 0.30     \\
	MCG+05-06-036    & $S^{*}$ & 41.77   & 42.68   & 0.78     & -162.46 & 0.01     \\
	MCG+07-23-019    & M                         & 42.03   & 42.74   & 0.39     & -84.19  & 0.10     \\
	MCG+08-11-002    & M                         & 41.33   & 42.33   & 0.81     & 69.80   & 0.28     \\
	MCG+08-18-013    & PM                         & 41.65   & 42.48   & 0.85     & -45.20  & 0.27     \\
	MCG+12-02-001    & M                         & 42.39   & 42.83   & 0.63     & 56.78   & 0.26     \\
	MCG-02-01-051    & PM                         & 42.42   & 42.82   & 0.69     & -49.62  & 0.50     \\
	MCG-02-01-052    & PM                         & 42.15   & 42.73   & 0.67     & -25.88  & 0.60     \\
	MCG-02-33-098    & M                         & 41.91   & 42.53   & 0.90     & -58.60  & 0.50     \\
	MCG-03-04-014    & PM                         & 42.30   & 42.85   & 0.81     & 20.84   & 0.25     \\
	MCG-03-34-064    & E                         & 42.51   & 42.93   & 0.82     & -54.02  & 0.39     \\
	MCG01-60-022     & LM                        & 42.11   & 42.59   & 0.74     & -87.54  & 0.27     \\
	Mrk231           & M                         & 43.29   & 43.99   & 0.79     & -3.05   & 0.28     \\
	Mrk331           & PM                         & 42.18   & 42.72   & 0.88     & -158.85 & 0.21     \\
	NGC 0317B        & M                         & 41.91   & 42.37   & 0.83     & 45.03   & 0.38     \\
	NGC 0838         & LM                        & 41.84   & 42.32   & 0.79     & 88.24   & 0.32     \\
	NGC 0958         & PM                         & 42.30   & 42.56   & 0.45     & -3.49   & 0.53     \\
	NGC 0992         & M                         & 42.09   & 42.44   &0.58     & -11.33  & 0.29     \\
	NGC 1068         & S                         & 42.89   & 43.10   & 0.67     & 53.14   & 0.05     \\
	NGC 1275         & $M^{*}$  & 42.94   & 43.14   & 0.74     & 80.00   & 0.10     \\
	NGC 1614         & M                         & 42.45   & 43.06   & 0.90     & 34.69   & 0.21     \\
	NGC 1797         & LM                        & 41.69   & 42.28   & 0.80     & 11.27   & 0.35     \\
	NGC 1961         & M                         & 42.35   & 42.50   & 0.46     & -81.00  & 0.50     \\
	NGC 2146         & M                         & 42.14   & 42.46   & 0.75     & 48.34   & 0.62     \\
	NGC 23           & S                         & 42.13   & 42.44   & 0.72     & 12.80   & 0.28     \\
	NGC 232          & PM                         & 41.38   & 42.42   & 0.82     & -5.94   & 0.23     \\
	NGC 2342         & M                         & 42.60   & 42.81   & 0.56     & -79.46  & 0.29     \\
	NGC 2388         & S                        & 41.77   & 42.41   & 0.77     & -74.63  & 0.01     \\
	NGC 2623         & M                         & 41.55   & 42.56   & 0.73     & -34.12  & 0.46     \\
	NGC 3110         & $S^{*}$  & 42.26   & 42.56   & 0.56     & 7.65    & 0.47     \\
	NGC 3221         & S                         & 41.66   & 42.15   & 0.70     & 10.05   & 0.71     \\
	NGC 34           & M                         & 41.65   & 42.68   & 0.76     & -34.23  & 0.33     \\
	NGC 3690(E)      & M                         & 41.83   & 43.24   & 0.78     & 83.92   & 0.10     \\
	NGC 3690(W)      & M                         & 42.17   & 43.26   & 0.77     & -64.14  & 0.10     \\
	NGC 4194         & M                         & 41.85   & 42.46   & 0.82     & 20.00   & 0.10     \\
	NGC 4418         & S                        & 40.34   & 42.47   & 0.53     & -63.71  & 0.53     \\
	NGC 4922         & M                         & 41.38   & 42.68   & 0.83     & 25.52   & 0.33     \\
	NGC 5104         & $LM^{*}$ & 41.82   & 42.34   & 0.60     & 11.41   & 0.48     \\
	NGC 5135         & S                         & 42.07   & 42.49   & 0.78     & -18.82  & 0.23     \\
	NGC 5256         & M                         & 42.70   & 42.97   & 0.38     & 41.98   & 0.39     \\
	NGC 5257         & PM                         & 42.48   & 42.84   & 0.55     & 45.00   & 0.20     \\
	NGC 5258         & PM                         & 42.13   & 42.72   & 0.74     & -41.19  & 0.42     \\
	NGC 5331         & PM                         & 42.06   & 42.66   & 0.95     & 58.42   & 0.50     \\
	NGC 5394         & M                         & 41.56   & 42.19   & 0.84     & 18.60   & 0.61     \\
	NGC 5395         & M                         & 41.92   & 42.31   & 0.44     & 4.50    & 0.60     \\
	NGC 5653         & S                         & 42.06   & 42.37   & 0.76     & 38.35   & 0.19     \\
	NGC 5734         & LM                        & 41.76   & 42.13   & 0.54     & -45.11  & 0.11     \\
	NGC 5936         & $S^{*}$  & 42.09   & 42.46   & 0.77     & -24.30  & 0.30     \\
	NGC 5990         & M                         & 41.98   & 42.41   & 0.84     & 66.79   & 0.32     \\
	NGC 6052         & M                         & 42.66   & 42.76   & 0.49     & -6.63   & 0.36     \\
	NGC 6090         & M                         & 42.72   & 43.03   & 0.82     & -84.92  & 0.11     \\
	NGC 6240         & M                         & 42.57   & 43.18   & 0.79     & -23.28  & 0.50     \\
	NGC 6286         & PM                         & 41.87   & 42.28   & 0.58     & -22.67  & 0.04     \\
	NGC 6621         & PM                         & 42.13   & 42.55   & 0.84     & 40.00   & 0.30     \\
	NGC 6670(E)      & M                         & 42.03   & 42.74   & 0.68     & 57.03   & 0.38     \\
	NGC 6670(W)      & M                         & 42.06   & 42.74   & 0.49     & -75.00  & 0.30     \\
	NGC 6701         & S                         & 42.04   & 42.38   & 0.68     & 37.64   & 0.04     \\
	NGC 6786         & PM                         & 42.38   & 42.84   & 0.65     & -43.87  & 0.38     \\
	NGC 6907         & PM                         & 41.83   & 42.27   & 0.90     & 81.34   & 0.47     \\
	NGC 6926         & M                         & 42.42   & 42.67   & 0.52     & -20.00  & 0.40     \\
	NGC 695          & M                         & 42.57   & 42.89   & 0.55     & 4.00    & 0.21     \\
	NGC 7469         & M                         & 42.81   & 43.15   & 0.84     & -72.32  & 0.23     \\
	NGC 7591         & PM                         & 41.45   & 42.28   & 0.90     & 23.80   & 0.50     \\
	NGC 7592(E)      & M                         & 42.24   & 42.64   & 0.60     & -13.00  & 0.40     \\
	NGC 7592(W)      & M                         & 41.83   & 42.52   & 0.78     & -15.93  & 0.36     \\
	NGC 7674         & PM                         & 42.44   & 43.01   & 0.76     & 10.15   & 0.10     \\
	NGC 7679         & LM                        & 42.45   & 42.64   & 0.74     & -175.80 & 0.06     \\
	NGC 7752         & PM                         & 42.15   & 42.41   & 0.73     & -76.70  & 0.53     \\
	NGC 7753         & PM                         & 41.97   & 42.31   & 0.29     & -48.87  & 0.38     \\
	NGC 7771         & PM                         & 41.89   & 42.46   & 0.72     & -72.83  & 0.67     \\
	NGC 828          & M                         & 42.33   & 42.57   & 0.77     & -75.00  & 0.30     \\
	NGC 877          & PM                         & 41.98   & 42.31   & 0.42     & 25.26   & 0.33     \\
	UGC 01385        & S                         & 41.93   & 42.39   & 0.92     & 25.03   & 0.20     \\
	UGC 03410        & PM                         & 41.74   & 42.14   & 0.74     & 58.57   & 0.74     \\
	UGC 04881        & M                         & 41.94   & 42.75   & 0.54     & -56.23  & 0.58     \\
	UGC 08387        & M                         & 41.32   & 42.65   & 0.81     & 31.01   & 0.51     \\
	UGC 08739        & PM                         & 41.46   & 41.99   & 0.73     & 58.15   & 0.71     \\
	UGC 11041        & $S^{*}$ & 41.99   & 42.30   & 0.62     & -46.16  & 0.33     \\
	UGC 12150        & S                         & 41.62   & 42.34   & 0.90     & -44.58  & 0.12     \\
	UGC 1845         & S                        & 41.54   & 42.20   & 0.82     & 34.59   & 0.35     \\
	UGC 2238         & M                         & 41.64   & 42.25   & 0.95     & 14.44   & 0.38     \\
	UGC 2369         & M                         & 42.03   & 42.97   & 0.89     & -86.15  & 0.27     \\
	UGC 2608         & M                         & 42.42   & 42.79   & 0.75     & -62.01  & 0.24     \\
	UGC 2982         & LM                        & 42.17   & 42.42   & 0.59     & 75.09   & 0.42     \\
	UGC 3094         & S                         & 42.20   & 42.60   & 0.60     & 5.00    & 0.60     \\
	UGC 3351         & PM                         & 41.59   & 42.13   & 0.67     & 5.79    & 0.42     \\
	UGC 3608         & M                         & 42.39   & 42.72   & 0.74     & -35.39  & 0.31     \\
	UGC 5101         & M                         & 41.81   & 42.92   & 0.73     & -72.24  & 0.41     \\
	VII ZW 031       & $E^{*}$  & 42.45   & 43.06   & 0.74     & 34.00   & 0.10     \\
	VV250a           & PM                         & 42.21   & 43.05   & 0.82     & 61.24   & 0.49     \\
	VV340a           & PM                         & 42.19   & 42.59   & 0.36     & -75.00  & 0.30     \\
	VV705            & M                         & 42.03   & 43.12   & 0.95     & 39.98   & 0.25      \\
	
	\hline  
	\hline
	\label{table:Ha}
\end{longtable*}

\clearpage

\appendix
\renewcommand{\appendixname}{Appendix~\Alph{section}}
	\section{R and $H\alpha$ images}

 \begin{flushleft}	
	
      \normalsize{By the way, we also show the R-band and continuum-subtracted $H\alpha$ images for each object in $appendix$ $A$, in order of object name. The solid line on the R-band images represent $10^{\prime \prime}$ and the object name is noted in the top center of images.} 
 \end{flushleft}	
	
	
	\begin{figure*}[!htb]
		\begin{center}	
			\includegraphics[page=1,scale=0.45]{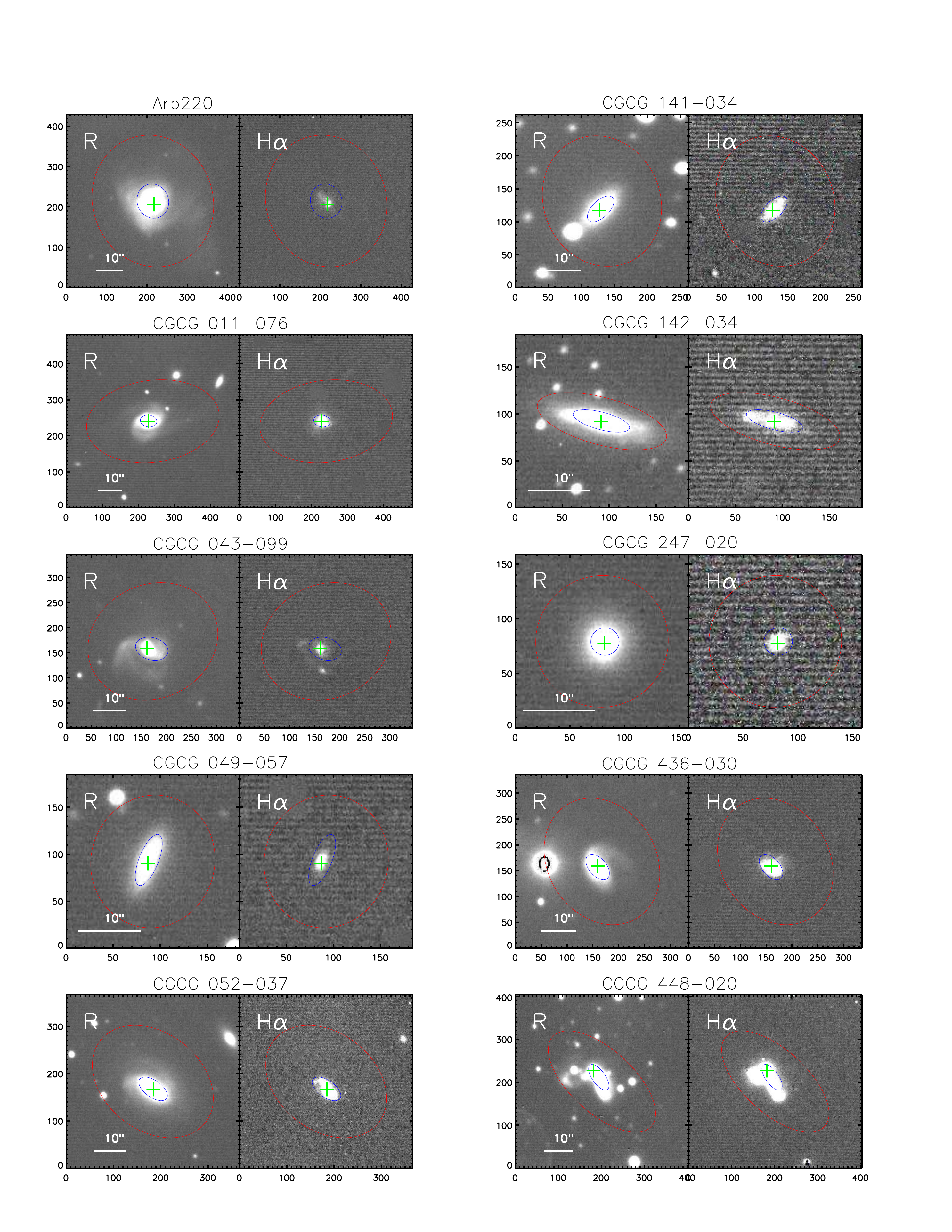}   
			\caption{ R-band image (left) and continuum-subtracted $H\alpha$ image (right). The images are listed in order of object name. The solid line on the R-band images represent $10^{\prime \prime}$. The other symbols on $R$-band and continuum-subtracted $H\alpha$ images are as the same as Figure ~\ref{fig:sub_Ha}. The object name is noted in the top center of images. } 
		\end{center}
	\label{fig:R_Ha} 
	\end{figure*}	
	\clearpage
	
	\includegraphics[page=2,scale=0.51]{counter.pdf}    
	\center{\normalsize{Figure 16. -continued}}
	\clearpage

	\includegraphics[page=3,scale=0.51]{counter.pdf} 
	\center{\normalsize{Figure 16. -continued}}
	\clearpage 
	
	\includegraphics[page=4,scale=0.51]{counter.pdf}
	\center{\normalsize{Figure 16. -continued}}
	\clearpage
	
	\includegraphics[page=5,scale=0.51]{counter.pdf}
	\center{\normalsize{Figure 16. -continued}}
	\clearpage
	
	\includegraphics[page=6,scale=0.51]{counter.pdf}
	\center{\normalsize{Figure 16. -continued}}
	\clearpage
	
	\includegraphics[page=7,scale=0.51]{counter.pdf}
	\center{\normalsize{Figure 16. -continued}}
	\clearpage
	
	\includegraphics[page=8,scale=0.51]{counter.pdf}
	\center{\normalsize{Figure 16. -continued}}
	\clearpage
	
	\includegraphics[page=9,scale=0.51]{counter.pdf}
	\center{\normalsize{Figure 16. -continued}}
	\clearpage
	
	\includegraphics[page=10,scale=0.51]{counter.pdf}
	\center{\normalsize{Figure 16. -continued}}
	\clearpage
	
	\includegraphics[page=11,scale=0.51]{counter.pdf}
	\center{\normalsize{Figure 16. -continued}}
	\clearpage
	
	\includegraphics[page=12,scale=0.51]{counter.pdf}
	\center{\normalsize{Figure 16. -continued}}
	\clearpage
	
	\includegraphics[page=13,scale=0.51]{counter.pdf}
	\center{\normalsize{Figure 16. -continued}}
	\clearpage
	
	\includegraphics[page=14,scale=0.51]{counter.pdf}
	\center{\normalsize{Figure 16. -continued}}
	\clearpage
	
	\includegraphics[page=15,scale=0.51]{counter.pdf}
	\center{\normalsize{Figure 16. -continued}}
	\clearpage
	
	\includegraphics[page=16,scale=0.51]{counter.pdf}
	\center{\normalsize{Figure 16. -end}}
	\clearpage

\section{Controversial source of morphological classification}
\begin{figure*}[!htb]
		\begin{center}	
			\includegraphics[page=1,scale=0.6]{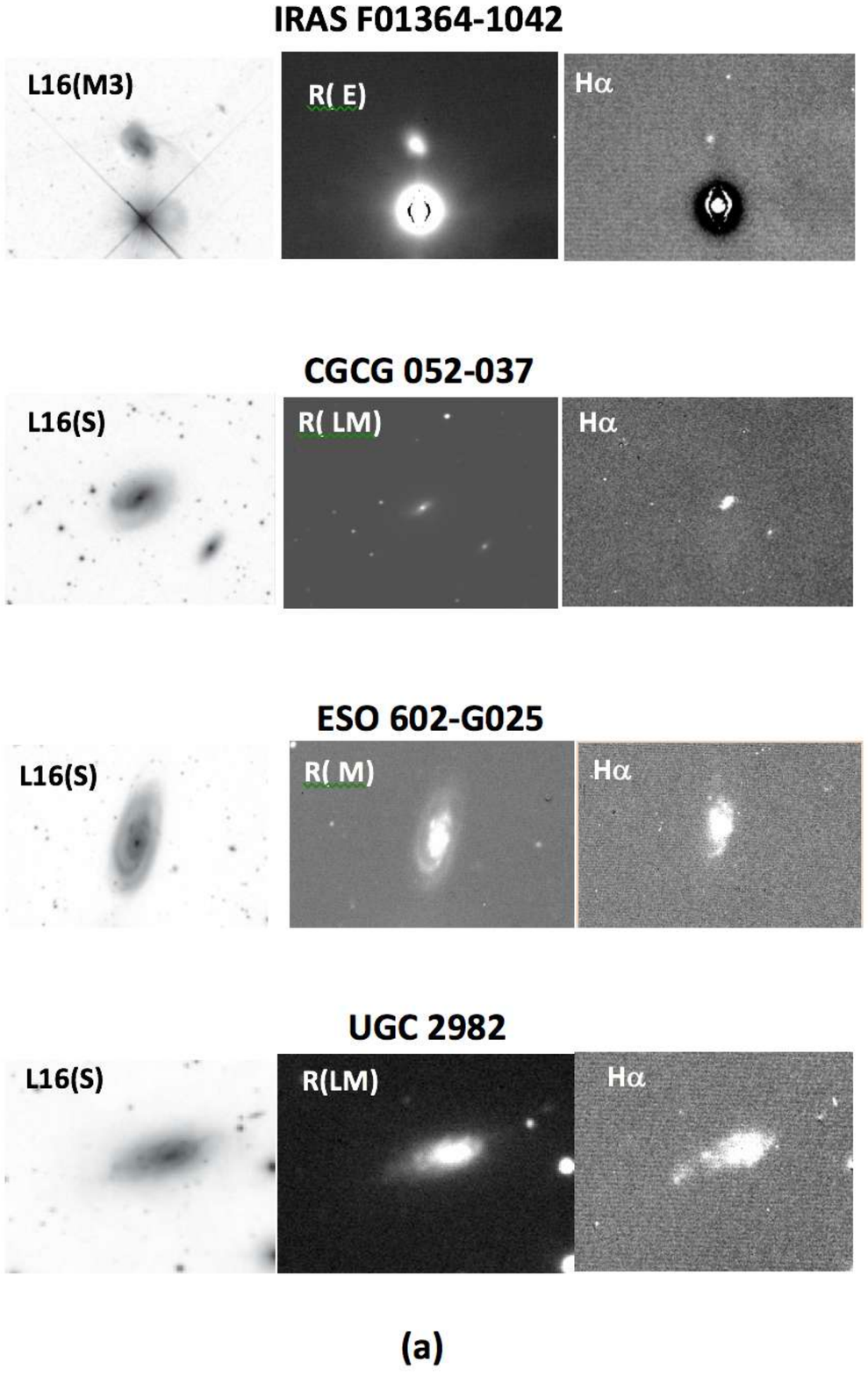}   
			\caption{The details and justifications for the classification of objects which require a change more than a single stage when compared with L16 and K13. The panel (a) lists the objects compared to L16, and the panel (b) lists the objects compared to K13. The left image in each row is derived from L16 (IRAS 20351+2521 and IRAS F17138-1017 don't have the images from L16), and the classification type made by L16 or K13 are marked. The middle image is the R-band image in this work, and the classification type made by this work are marked. The right image is the $H\alpha$ image in this work. } 
		\end{center}
	\label{fig:com} 
	\end{figure*}

 \begin{flushleft}	
\normalsize{\textbf{    In this section, we provide further details and discussion for the classification of 8 objects which are not exactly matched with L16 or K13. We will state why our classification differs from L16 or K13, and the reasons for our classification. The images for these objects are listed in Figure ~\ref{fig:com}}

~\\
\textbf{IRAS F01364-1042} classified as E in our work based on the detection of a single compact nucleus and there are no signs of tides can be distinguished in our image. This object is classified as  M3 in L16 based on disturbed disk with small projected nuclear separation. The difference between this two classification is due to difference in resolution, and ultimately, we retain our original decision for consistency among our sample.\\

~\\
\textbf{CGCG 052-037} classified as LM in our work based on it seems to have some tidal structure in this galaxy.  This object is classified as s in L16 based on it's appearance of a single object with no clear sign of interaction. The difference in classification for this objects is more due to subjective.\\

~\\
\textbf{ESO 602-G025} classified as M in our work based on it seems have some interaction structure which may be due to a minor merge with a small galaxy. This object is classified as s in L16 based on they think there is no clear sign of an interaction. The difference in classification for this objects is more due to subjective.

~\\
\textbf{UGC2982} classified as LM in our work based on it's tidal structure. This objects is classified as s in L16. Combined with the $H\alpha$ image, we believe that this object is more likely to be in an interaction.

~\\
\textbf{IC5298} classified as LM in our work based on single nucleus with faint tidal tail. This objects is classified as 0 (single undisturbed galaxy, shows no signs of tidal interaction) in K13.  In L16, this object is classified as m (minor merge). They think there is a small companion in SW and they are connected together with a tidal tail.  In a way, our result is consistent with L16.

~\\
\textbf{IRAS 20351+2521} classified as M in our work based on it's clearly disturbed disk. L16 don't contain this object and K13 classify this object as 0. 

~\\
\textbf{IRAS F17138-1017} This object is not contain in the work of L16. We classify this object as M based on it's disturbed disk. But K13 classify this object as 0. The difference in classification for this objects is more due to subjective.

~\\
\textbf{NGC695} classified as M in our work based on it's disturbed disk. This objects is classified as 0 in K13.  In L16, this object is classified as m (minor merge). They think there is a minor companion NW of the main galaxy along with appearance of a tidal perturbation. In a way, our result is consistent with L16.
}
 \end{flushleft}	
	
	
	\clearpage
	
	\includegraphics[page=1,scale=0.6]{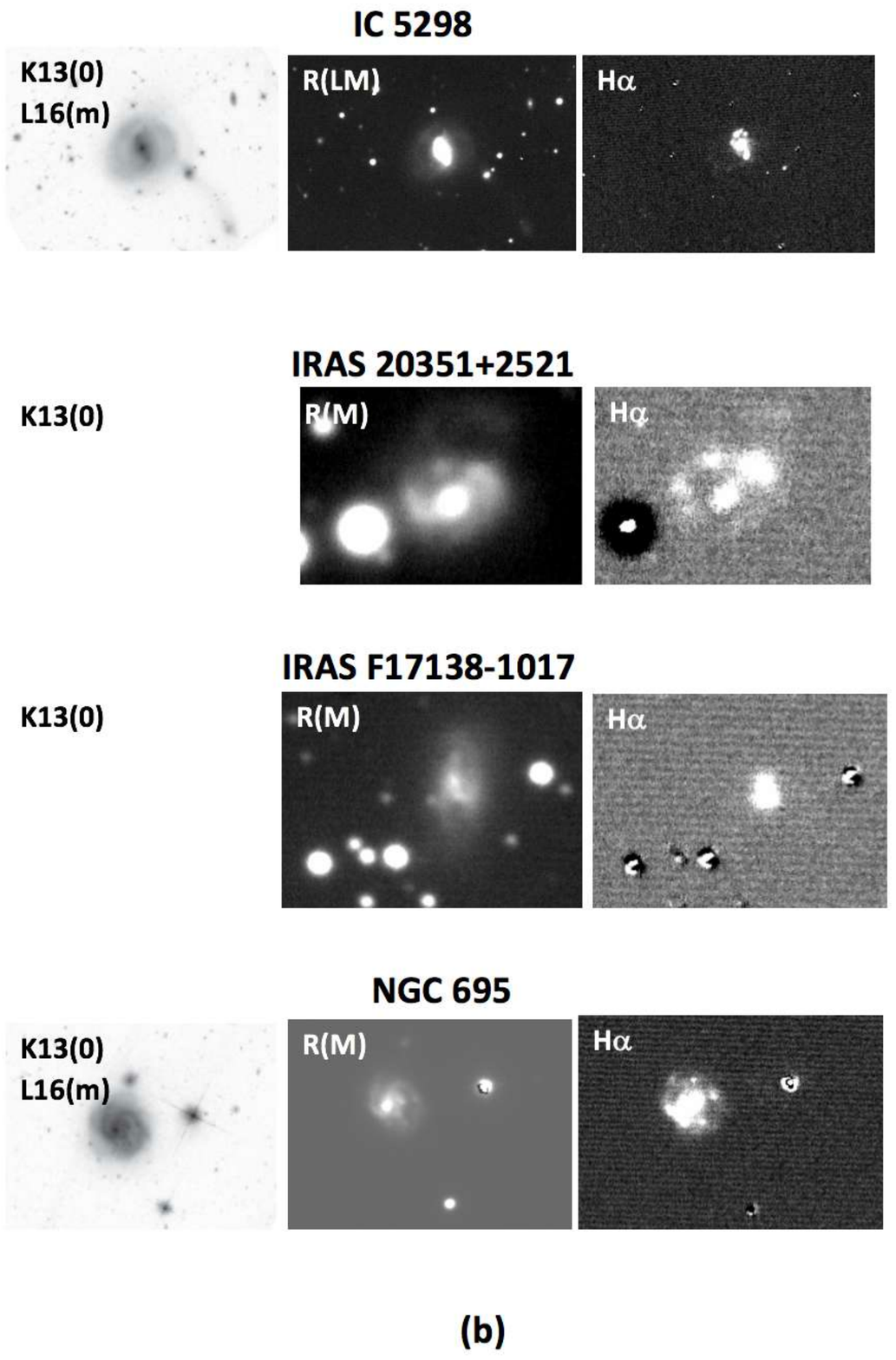}    
	\center{\normalsize{Figure 17. -continued}}
	\clearpage


\end{document}